\documentclass{aa}  

\usepackage{graphicx}
\usepackage{txfonts}
\usepackage{lipsum}
\usepackage{subcaption} 
\usepackage{lscape}    
\usepackage{placeins}          
\usepackage{bm}
\usepackage[normalem]{ulem}
\usepackage[dvipsnames]{xcolor}

\usepackage{orcidlink}

\newcommand{\dpar}{d_{\parallel}}
\newcommand{\dperp}{d_{\perp}}
\newcommand{\dfl}{d_{FL}}
\newcommand{\Dpar}{D_{\parallel}}
\newcommand{\Dperp}{D_{\perp}}
\newcommand{\Dpars}{D_{\parallel}^*}
\newcommand{\Dperps}{D_{\perp}^*}

\newcommand{\lpa}{\lambda_\parallel}
\newcommand{\lpe}{\lambda_\perp}

\newcommand{\kbold}{\bm{k}}

\newcommand{\dd}{\mathrm{d}}
\newcommand{\ii}{\mathrm{i}}
\newcommand{\e}{\mathrm{e}}
\newcommand{\vpar}{v_{\parallel}}
\newcommand{\vperp}{v_{\perp}}

\newcommand{\zhat}{\hat{\bm z}}
\newcommand{\xhat}{\hat{\bm x}}
\newcommand{\yhat}{\hat{\bm y}}
\newcommand{\bhat}{\hat{\bm b}}
\newcommand{\R}{\bm R}
\newcommand{\vrho}{\bm\rho}
\newcommand{\kvec}{\bm k}

\newcommand{\avg}[1]{\left\langle #1 \right\rangle}
\newcommand{\gyroavg}[1]{\left\langle #1 \right\rangle_{\phi}}

\usepackage{hyperref}
\hypersetup{
    pdftitle           = {Perpendicular transport in isotropic magnetic turbulence: The role of stochastic drifts},
    pdfauthor          = {F./ Pucci et al.}
    pdfsubject         = {UHECR},   
    pdfkeywords        = {Cosmic Rays}, 
    pdfpagemode        = {UseOutlines},  
    pdfcenterwindow    = {true},
    pdfdisplaydoctitle = {true},
    pdffitwindow       = {true},
    pdfpagelayout      = {SinglePage},
    pdfstartview       = {FitH},         
    pdfstartpage       = {1},
    pdflang            = {en-US},
    colorlinks         = true,       
    citecolor          = magenta,
    linkcolor          = magenta,        
    filecolor          = blue,   
    urlcolor           = blue,        
    breaklinks         = true,
    bookmarksopen      = true,
    bookmarksnumbered  = true
}

\begin{document}

   \title{Perpendicular transport in isotropic magnetic turbulence}
   \subtitle{The role of stochastic drifts}


   \author{F. Pucci\inst{1}\fnmsep\thanks{These authors contributed equally and should be considered joint first author.}\fnmsep\thanks{Corresponding author: francesco.pucci@istp.cnr.it}\orcidlink{0000-0002-5272-5404}, 
   P. Blasi\inst{2,3}\fnmsep\textsuperscript{$\star$}\orcidlink{0000-0003-2480-599X}, 
   O. Pezzi\inst{1}\orcidlink{0000-0002-7638-1706}, C. Evoli\inst{2,3}\orcidlink{0000-0002-6023-5253}, and F. Malara\inst{4}\orcidlink{0000-0002-5554-8765}
        }

   \institute{Consiglio Nazionale delle Ricerche, Istituto per la Scienza e Tecnologia dei Plasmi, Via Amendola 122/D, 70126, Bari (BA), Italy
              \email{francesco.pucci@istp.cnr.it}
              \and Gran Sasso Science Institute (GSSI), Viale Francesco Crispi 7, 67100 L’Aquila, Italy 
              \and INFN-Laboratori Nazionali del Gran Sasso (LNGS), via G. Acitelli 22, 67100 Assergi (AQ), Italy  
              \and Università della Calabria, Dipartimento di Fisica, 87036, Arcavacata di Rende, Italy         
             }
             
   \date{Received ...}

 
  \abstract
{In the context of cosmic ray propagation in a turbulent magnetic field, the transport perpendicular to the mean field has an important role in a variety of situations of physical interest. Numerical simulations of test particles in isotropic turbulence with a guide field have shown that the parallel and perpendicular diffusion coefficients scale differently with energy over a wide range of energies for $r_L<L_c$, so that the corresponding ratio $D_\perp/D_\parallel$ grows with energy. Some attempts have been made to address this phenomenon, but there is at present no consensus theoretical understanding.}
{Here we use test particle numerical simulations of particle transport in synthetic turbulence over a wider range of energies to establish a general trend of the perpendicular diffusion coefficient, and we provide a theoretical description of our findings based on an extension of the non-linear guiding center theory that includes stochastic drifts.} 
{The numerical simulations of isotropic turbulence are based on a wavelet method that allows us to reproduce turbulence over $\sim 6-7$ orders of magnitude in wavenumber. The latter is reflected in a correspondingly wide energy range for test particles. The theoretical part is a novel description of perpendicular transport that starts from auto-correlation functions of turbulent magnetic fields to infer how particles diffuse perpendicular to the field lines due to a combination of field line random walk and random drift motion of the guiding center.}
{We demonstrate that, for diverse levels of turbulence, the perpendicular diffusion coefficient has the same energy dependence as that of the parallel one only for energies corresponding to $r_L\lesssim 10^{-2}L_c$, where $r_L$ is the particle Larmor radius and $L_c$ is the coherence length of the turbulence. At higher energies, the perpendicular diffusion coefficient acquires a stronger energy dependence. Our theoretical approach suggests that this phenomenon is likely attributable to stochastic drifts that particles experience in the direction perpendicular to the guide field when a component of the turbulent field is parallel to it.}
{}


   \maketitle
\nolinenumbers

\section{Introduction}

Investigations of the origin of Galactic cosmic rays (CRs) heavily rely on a description of the propagation of charged particles in turbulent magnetic fields. In phenomenological approaches to Galactic transport, the complexity of the problem is summarised through the introduction of a scalar diffusion coefficient, which is typically a function of energy and, in some cases, changes with the location in the Galaxy (see \citealt{Strong:2007nh,Evoli:2012ha,Gabici:2019jvz} and references therein). 

On the other hand, in the presence of a regular large scale magnetic field ${\mathbf B_0}$, and low or moderate levels of magnetic turbulence, particle diffusion behaves in very different ways in the direction{\bf{s}} parallel and perpendicular to ${\mathbf B_0}$, with $D_\perp\ll D_\parallel$ in general, being $D_\parallel$ and $D_\perp$ the diffusion coefficients in the directions parallel and perpendicular to ${\mathbf B_0}$ \cite[]{casse2001transport,DeMarco2007,dundovic2020novel,giacinti2018reconciling}. 
The problem of perpendicular transport has been discussed in many contexts and with numerous assumptions on the type of turbulence, more prominently using slab+2D \cite[]{ShalchiBook2009,Shalchi2019,shalchi2021perpendicular}.
The test particle simulations of transport in isotropic synthetic turbulence by \citet{DeMarco2007} suggested for the first time that the energy dependence of $D_\perp$ was stronger than that of $D_\parallel$, different from results obtained with a slab+2D turbulence realization that showed an opposite trend \citep{Shalchietal2004}.

This finding was investigated and confirmed by several subsequent pieces of work \citep{husseinetal2015,giacinti2018reconciling,dundovic2020novel}. In all these cases, simulations were carried out by generating Fourier modes in a three-dimensional box and propagating test particles using periodic boundary conditions. The high computational cost of these simulations implies that only a relatively narrow range of wavenumbers (and hence particle energies) can be investigated, typically less than 2 decades below the turbulence coherence scale $L_{c}$ and about 1 decade above. In other words, in terms of Larmor radius, these simulations cover $10^{-2}\lesssim r_L/L_c\lesssim 10$. 
Recently, thanks to the use of graphics processing units (GPUs) and nested bozes techniques, \citet{kuhlen2025simulations} were able to extend such an energy range by a few orders of magnitude.



The energy dependence of the perpendicular transport was investigated in detail by \cite{dundovic2020novel}, which demonstrated how the stronger energy dependence of $D_\perp$ with respect to $D_\parallel$ disappears in models in which the turbulence is 2D or 2D+slab, without a component of the turbulent field along the ordered magnetic field (namely cases in which $\delta B_z = 0$ and the guide field is directed along z). An even more extended study of particle scattering in different models of turbulence, including slab+2D and Goldreich and Sridhar models \cite[]{GS1994,GS1995}, had been previously carried out by \citet{husseinetal2015}, with qualitatively similar results.

\citet{dundovic2020novel} also developed a calculation of $D_\perp$ based on non-linear guiding center theory (NLGCT) \cite[]{matthaeus2003nonlinear} (see also \citealt{ShalchiBook2009} for a review), and proved that for a wide range of cases, for isotropic turbulence, the ratio $D_\perp/D_\parallel$ returned by the theoretical calculations is constant at low energies and decreases with energy in the energy range $10^{-2}\lesssim r_L/L_c\lesssim 1$. This work proved that whatever causes the anomalous energy dependence of $D_\perp (E)$ is not due to standard diffusion, and, perhaps more importantly, it is peculiar to 3D isotropic turbulence ($\delta B_z\neq 0$). 

More recently, \cite{kuhlen2025simulations,kuhlen2025models} carried out similar numerical simulations but used nested boxes techniques, which allowed them to extend their conclusions to a broader energy range for the propagated test particles. These simulations showed that at sufficiently low energies the flat $D_\perp/D_\parallel$ ratio as a function of energy is recovered, while confirming a growing ratio for $10^{-2}\lesssim r_L/L_c\lesssim 1$, as previously found by \cite{DeMarco2007,giacinti2018reconciling,husseinetal2015,dundovic2020novel}. 
It is worth stressing that all this work concurs in deducing that there is no anomaly in the description of parallel transport, whose energy dependence seems to be well predicted by standard considerations based on resonant scattering: 
the slope of $D_\parallel (E)$ as a function of energy is exactly as expected, 1/3 (1/2) for a Kolmogorov (Kraichanan) spectrum at $r_L/L_c\lesssim 1$ and $\sim 2$ for $r_L/L_c\gg 1$ (independent of the spectrum of the adopted turbulence). The latter implies that the unexpected behavior is attributable solely to perpendicular transport. 

\cite{kuhlen2025simulations,kuhlen2025models} suggested that the anomalous behavior of the ratio $D_\perp/D_\parallel$ may be due to the onset of a subdiffusive phase in the random walk of magnetic field lines and presented a semi-analytic model in support of this interpretation of their numerical results. The subdiffusive phase is, in fact, found especially for low levels of turbulence $\zeta=\delta B/B_0\lesssim 1$, while it becomes less prominent or even disappears for $\zeta\gtrsim 1$. For $\zeta\sim 1$ there is no clear subdiffusion of the field lines, and yet the ratio $D_\perp/D_\parallel$ increases with energy for $10^{-2}\lesssim r_L/L_c\lesssim 1$, which led us to question the interpretation of \cite{kuhlen2025simulations,kuhlen2025models}, or at least to question that the subdiffusive behavior of the field line random walk may be the sole responsible for the observed energy dependence of perpendicular transport. 

The purpose of the present article is twofold: on one hand, we present the results of new test particle simulations in which turbulence is produced by using a wavelet method, first introduced by \cite{malara2016fast}, which allows a fast, numerically agile method to simulate transport of test particles in an energy range $5\times 10^{-5}\lesssim r_L/L_c\lesssim 10$, somewhat wider than reached by \cite{kuhlen2025simulations}. This part confirms the previous result \cite[]{kuhlen2025simulations} that there is a change in perpendicular transport in the energy region $10^{-2}\lesssim r_L/L_c\lesssim 1$, while for $r_L/L_c\lesssim 10^{-2}$, the energy dependence of $D_\parallel$ and $D_\perp$ seems to be the same (these conclusions seem to depend weakly on the level of turbulence). On the other hand, we suggest a different physical interpretation of this result, due to random drifts associated with fluctuations of the turbulent magnetic field (mainly) in the z-direction. Specifically, we develop a nonlinear guiding-center approach to the newly identified phenomenon of random drifts and show that the calculations reproduce the energy dependence of $D_\perp$ for the turbulence levels investigated here. However, the strength of this conclusion depends on certain details of the closure adopted in our theoretical approach. 

It is worth stressing that the regime of three-dimensional isotropic turbulence is probably to be associated with something close to fast magnetosonic turbulence, in which case the turbulent cascade is expected to develop quasi-isotropically. For Alfv\'enic turbulence, it is now accepted that the cascade is highly anisotropic \cite[]{GS1994,GS1995} and primarily develops in the perpendicular direction in $k$-space. This leads to a suppression of the role of resonances and, correspondingly, a very weak particle scattering rate. It has been recently proposed that occasional interactions with intermittent structures may dominate particle scattering in realistic MHD turbulence \cite[]{Lemoine2023,Kempski2023,Kempski2025}, rather than resonant scattering. These effects are not testable in models of synthetic turbulence such as the one used here. 

The article is structured as follows: in Section \ref{sec:simula}, we describe our simulation approach, while the findings of these simulations, in terms of parallel and perpendicular transport at different levels of turbulence, are presented in Section \ref{sec:results}. In Section \ref{sec:simple}, we discuss an analytic approach, presented in \cite{Shalchi2019,kuhlen2025models}, that allows us to infer simple scalings for standard perpendicular diffusion in terms of the random walk of magnetic field lines and parallel particle transport. In Section \ref{sec:drifts}, we discuss the details of the theoretical approach we propose, based on a generalization of NLGCT that describes both standard perpendicular diffusion and the contribution due to stochastic drifts. We compare the theoretical results with those of our numerical simulations in Section \ref{sec:TS}. We draw our conclusions in Section \ref{sec:conclude}.


\section {The numerical method}
\label{sec:simula}
We performed test-particle simulations of cosmic-ray diffusion in isotropic magnetic turbulence plus a guide field. These consist of solving the relativistic equations of motion for a charged particle under the action of the Lorentz force, which in normalized units read 
\begin{align}
    \frac{d{\bm x}}{dt}&={\bm v} \label{eq1} \\
    \frac{d{\bm v}}{dt}&= {\bm v} \times {\bm B}. \label{eq2}
\end{align}
In Equations~(\ref{eq1})-(\ref{eq2}), the particle position ${\bm x}$ and all lengths are normalized to a typical length $L$; the magnetic field ${\bm B}$ is normalized to a typical value $B_0$; time is normalized to the gyration frequency $\Omega^{-1}=\frac{m\gamma}{qB_0c}$, where $q$, $m$, and $\gamma$ are the particle charge, mass, and Lorentz factor, respectively; particle velocity ${\bm v}$
is normalized to $V=L\Omega$. The equations of motion are solved numerically using the Boris algorithm \citep{qin2013why,pucci2016energetic}. 
We investigate cosmic-ray transport using test-particle simulations in a prescribed, magnetostatic turbulent magnetic field:
\begin{equation}
\bm{B}(\bm{r})=\bm{B}_0+\delta\bm{B}(\bm{r}),
\end{equation}
where $\bm{B}_0$ is a uniform mean field directed in the $z$ direction and the fluctuations $\delta\bm{B}(\bm{r})$ are generated with the fast synthetic-turbulence algorithm introduced by \citet{malara2016fast}.
The turbulent field is (i) three-dimensional and periodic, (ii) exactly solenoidal, and (iii) sufficiently regular (continuous derivatives) to allow accurate orbit integration without spurious numerical scattering. To satisfy $\nabla\cdot\bm{B}=0$ by construction, the model represents the fluctuating field as a superposition of localized eddies written in curl form,
\begin{equation}
\delta\bm{B}(\bm{r})=\sum_{m=m_I}^{N_s}\sum_{i,j,k}\delta\bm{B}^{(i,j,k;m)}(\bm{r}),
\quad
\delta\bm{B}^{(i,j,k;m)}(\bm{r})=\nabla\times\bm{A}^{(i,j,k;m)}(\bm{r}),
\end{equation}
so that each eddy is defined by a vector potential $\bm{A}^{(i,j,k;m)}$ with compact support.
The model emulates a turbulent cascade through a hierarchy of dyadic cells: at scale index $m$ the periodic domain $[0, L_x]\times[0, L_y]\times[0, L_z]$ is decomposed into $2^{3m}$ cells of size $\ell_{x,m}=2^{-m}L_x$ (and similarly for $y,z$), with a typical scale $\ell_m=2^{-m}L_0$, where $L_0=(L_xL_yL_z)^{1/3}$. In this work, we consider a cubic box with $L_x=L_y=L_z$; the index $m$ runs from $m_I$, which determines the typical size of the larger wavelet, and $N_s$, which is the total number of scales.
The indices $i,j,k$ uniquely identify a specific cell at scale $m$. Each cell hosts one eddy, but the eddy support is chosen to be larger than the cell itself (by a factor of 2 in each direction), so that neighboring eddies overlap; this prevents the artificial vanishing of fluctuations at cell boundaries and improves statistical homogeneity.

The eddy shape is prescribed using a smooth polynomial ``base function'' $F(t)$ that is nonzero inside the support and vanishes together with several derivatives at the boundary, guaranteeing a turbulent field that is continuous with high regularity.
The base functions are further modified by a nonlinear mapping controlled by random parameters $\phi_{x,y,z}$ (drawn in $[-1,1]$) that distort the eddy shape and further reduce residual lattice imprinting, while preserving the regularity of the construction.

To reproduce a desired inertial-range spectrum, eddy amplitudes follow a scale-dependent power law consistent with a prescribed spectral slope (e.g.\, Kolmogorov slope corresponds to the usual $h=1/3$ scaling for increments).
Beyond reproducing second-order statistics, the method can incorporate intermittency through a multiplicative cascade based on the classical $p$-model (see \citealt{malara2016fast} for further details). However, in this work, we will suppress intermittency ($p=0.5$), thus considering a case in which the probability distribution function of fluctuations is Gaussian and the kurtosis is 3 at all scales.

A central practical advantage for orbit calculations, which also belongs to other numerical methods (see, e.g., \citealt{Giacalone1999}), is the model’s \emph{fast, grid-free evaluation}. Although the total number of eddies grows exponentially with the number of scales $N_s$, the field at a given position $\bm{r}$ depends only on the finite set of eddies whose supports contain $\bm{r}$; accounting for overlap, there are eight contributing eddies per scale. Thus, the full field can be evaluated as
\begin{equation}
\delta\bm{B}(\bm{r})=\sum_{m=m_I}^{N_s}\sum_{\mu=1}^{8}\nabla\times\bm{A}^{(\mu;\bm{r};m)}(\bm{r}),
\end{equation}
so that the computational cost scales as $8N_s$ rather than with the total number of cells. Moreover, the random parameters defining each eddy are not stored. Instead, each cell is assigned an ``absolute address'' $\lambda(i,j,k;m)$, which is used as a deterministic seed for a random-number generator; by calling the generator a fixed number of times, the required parameters are reconstructed on the fly whenever that eddy is needed.
This eliminates the prohibitive memory demands that would arise from storing all eddies for large $N_s$, and it allows us to generate ensembles of statistically independent turbulence realizations simply by changing the sampling index used in the random-number routine.

Our synthetically generated turbulent field $\delta {\bf B}$ is isotropic. The latter means that i) the spectrum is isotropic in the 3D wavenumber space, ii) the root mean squares of each magnetic field component are equal, and $$\delta B = \sqrt{\delta B_x^2+\delta B_y^2+\delta B_z^2}=\sqrt{3}\delta B_x.$$ Our model contains three characteristic scales: the periodicity length $L_0$, the turbulence correlation length $L_c$, which is controlled by varying $m_I$, and the minimum turbulence scale $L_{min}$, which is controlled by $N_s$. In the simulations we present, we set $m_I=2$ and $N_s=23$, which results in $L_0/L_c = 20$ and $L_c/L_{min}=5\times10^7$. The magnetic energy spectrum in the range $1/L_c<k<1/L_{min}$ follows a Kolmogorov scaling $\propto k^{-5/3}$ as prescribed by the model. Such an extended dynamical range is unprecedented in the existing literature and is a direct outcome of our wavelet-based method. 

On the other hand, for wavenumbers $k<1/L_c$, the energy spectrum depends on the spectrum of the largest wavelet and scales as $\sim k^3$. This is an important point to keep in mind, for two reasons. Firstly, the condition of homogeneity of the turbulence would impose a low-$k$ behavior $\propto k^4$ \cite[]{Batchelor1982} (see also discussion in \citet{dundovic2020novel}). In this perspective and limited to the low-$k$ range, the numerical approach adopted here does not fully reproduce the properties of homogeneous turbulence. Notice, however, that this is a problem common to most numerical simulations. For instance, in \citet{dundovic2020novel}, the spectrum did not follow exactly the low-$k$ theoretical prediction $\propto k^4$ due to the limited numerical separation between the box size and the bend-over scale. Moreover, \citet{kuhlen2025simulations} used a power spectrum with a low-$k$ truncation at a minimum wavenumber $k_0$. Secondly, while the total power in the form of turbulence is fixed, its distribution in $ k$-space varies, leading to differences in the perpendicular transport of particles. We will discuss this point in more detail in Section \ref{sec:drifts}.

\section {Results of numerical simulations}
\label{sec:results}

We performed numerical simulations for three different values of the turbulence amplitude $\zeta=\delta B/B_0= 0.3,0.5,1.0$, where $B_0$ is the intensity of the guide-field. For each of these values, we span the energy range corresponding to $2\times10^{-4}\leq r_L/L_c\leq 20$, where $r_L=c/\Omega$ is the particle Larmor radius. 
For the sake of having some physical numbers in mind, if one considers a correlation length of the turbulence $L_c=10\mathrm{pc}$ and a magnetic field  $B=1\mu\mathrm{G}$, which are typical values for the Galactic interstellar medium \citep{beck2001,elmegreenandscalo2004,haverkornetal2004}, the particle energy range we simulate runs from $1.85\mathrm{TeV}$ to $185\mathrm{PeV}$, which mainly falls in the range relevant for high-energy Galactic CRs \citep{strongetal2007}. 

The time step in the Boris algorithm is fixed to $dt=2\pi/50\Omega^{-1}$. We computed the trajectories of $4096$ particles for each simulation. For each run at a given energy and $\zeta$, the particles are initialized with random positions, equal speed, and an isotropic distribution of the initial velocity directions. The simulations are conducted in parallel, using one computing core and one process to evolve a single particle.
Particle positions are saved at a fixed cadence in logarithmic scale to reduce the size of the output data. Particle trajectories are computed until a diffusive regime is statistically achieved in both the parallel and perpendicular directions with respect to the guide field.   


Now, we define a few useful quantities. The running parallel and perpendicular diffusion coefficients are defined as 
$$\dpar(t) = \frac12\frac{d\langle\Delta z^2\rangle}{dt}$$
and 
$$\dperp(t) = \frac12\frac{d\langle\Delta x^2\rangle}{dt}=\frac12\frac{d\langle\Delta y^2\rangle}{dt},$$ 
where the particle position displacements are $\Delta x(t) = x(t)-x(0)$, $\Delta y(t) = y(t)-y(0)$, $\Delta z(t) = z(t)-z(0)$ (being the particle position ${\bm x}=(x,y,z)$), and $\langle\rangle$ means average over all the particles. We also define 
\begin{equation}\label{Eq:Dpar}
\Dpar=\frac{\langle\Delta z^2\rangle}{2t}
\end{equation}
and 
\begin{equation}\label{Eq:Dperp}
\Dperp=\frac{\langle\Delta x^2\rangle}{2t}=\frac{\langle\Delta y^2\rangle}{2t}.
\end{equation}
The asymptotic diffusion coefficients are $\Dpars=\lim_{t\to\infty}\Dpar(t)=\lim_{t\to\infty}\dpar(t)$ and
$\Dperps=\lim_{t\to\infty}\Dperp(t)=\lim_{t\to\infty}\dperp(t)$. Both $d_{\parallel(\perp)}$ and $D_{\parallel(\perp)}$ saturate to the same value, although showing small differences in their time evolution: in particular,  $d_{\parallel(\perp)}$ saturates slightly earlier in time compared to $D_{\parallel(\perp)}$. The mean parallel and perpendicular free paths are then defined as $\lpa=3\Dpars/c$ and $\lpe=3\Dperps/c$. Finally, the magnetic field line diffusion coefficient is defined as $$ \dfl(s) = \frac12\frac{d\langle \Delta x^2_{FL} \rangle}{ds}=\frac12\frac{d\langle \Delta y^2_{FL} \rangle}{ds},$$ where $s$ is the field-aligned curvilinear coordinate.

\begin{figure}[h!]
 \centering
 \includegraphics[width=\hsize]{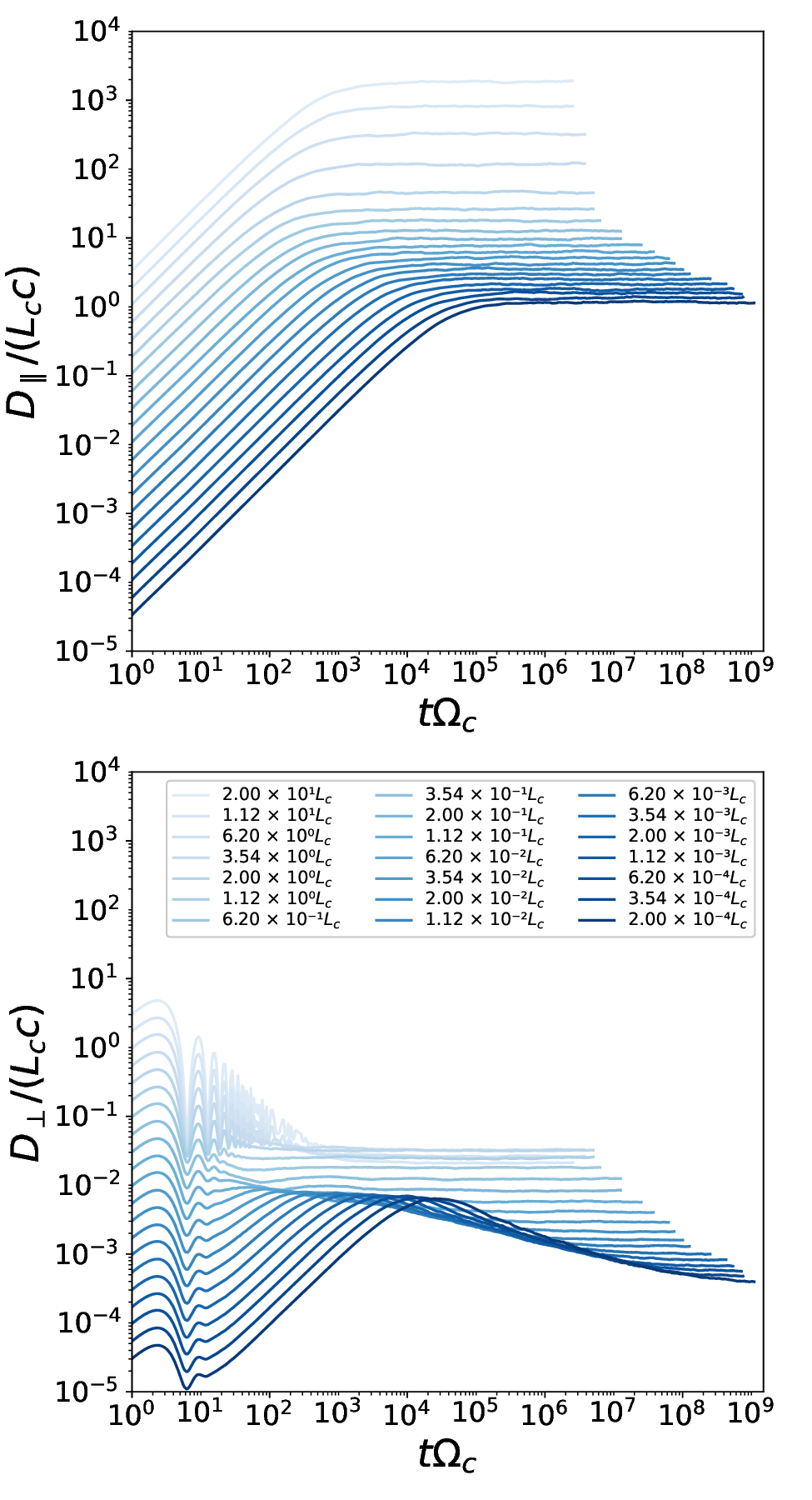}
   \caption{Parallel and perpendicular running diffusion coefficients as a function of time at different energies for $\zeta=0.3$. The values are computed from test-particle simulations as in Equations~(\ref{Eq:Dpar})-(\ref{Eq:Dperp}), averaging over particle trajectories.}
         \label{fig1}
 \end{figure}

Figure~\ref{fig1} shows the evolution of $\Dpar(t)$ and $\Dperp(t)$ at different energies for $\zeta=0.3$; the figure is meant to show that diffusion is achieved in both parallel and perpendicular transport at all the energies considered. The same applies to the less computationally challenging cases $\zeta=0.5$ and $\zeta=1.0$ (not shown here).
The evolution of diffusion coefficients follows the standard behavior. Parallel transport is ballistic until the parallel-scattering time is reached. After that, the diffusive regime begins. The parallel scattering time in units of the gyration time increases with decreasing particle energy. In the perpendicular direction, oscillations are observed at earlier times as a result of the Larmor gyration. 

\begin{figure}[h!]
 \centering
 \includegraphics[width=\hsize]{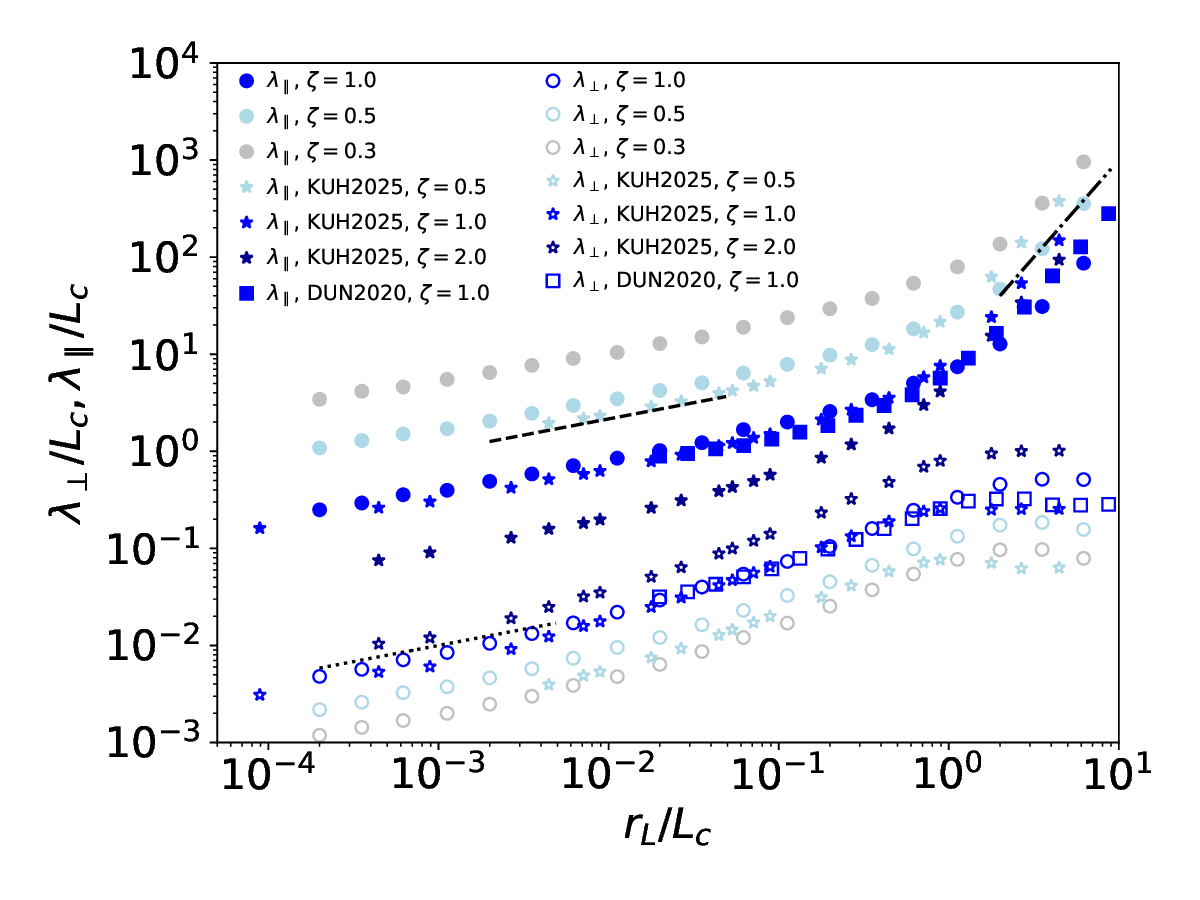}
   \caption{Parallel and perpendicular mean free path as a function of the particle energy. Different colours correspond to different turbulence levels: $\zeta=0.3$ (silver), $\zeta=0.5$ (light-blue), $\zeta=1.0$ (blue), $\zeta=2.0$ (darkblue). Different symbols correspond to this work (circles), \citet{kuhlen2025simulations} (KUH2025, stars), and \citet{dundovic2020novel} (DUN2020, squares). For reference, the dashed and dotted lines correspond to the $\propto(r_L/L_c)^{1/3}$ power-law scaling, while the dash-dotted lines correspond to the $\propto(r_L/L_c)^{2}$ power-law scaling at high energies.}
         \label{fig2}
 \end{figure}

In the perpendicular direction, oscillations are observed at early times as a result of the Larmor gyration and are more evident for longer times for particles having $r_L \gtrsim L_c$. At lower energies, $D_\perp(t)$ initially grows with time due to free parallel motion along the field line and field line ballistic transport. The evolution at later times is determined by the nature of the field line and parallel transport. Specifically, in our case we can assume that field lines become diffusive after particles have moved $s \sim L_c$ in the direction parallel to the mean field. Since the parallel mean free path $\lambda_\parallel$ is larger than $L_c$ for $\zeta=0.3$ (see Figure~\ref{fig2}), the field lines become diffusive before parallel scattering takes place, leading to the so-called field line random walk limit \citep{matthaeusetal1995}. The latter corresponds to the maximum value at which $D_\perp(t)$ saturates for $r_L/L_c\lesssim6\times10^{-2}$. We notice already that particles whose Larmor radius is smaller but close to $L_c$ ($r_L \lesssim L_c$) reach asymptotic values of $D_\perp$ that are larger than the field line random walk limit, suggesting that they are not exactly following field lines. We will discuss the origin of this behavior later in this work. At later times, or longer distances traveled by the particles along the mean field, particles start scattering in the parallel direction, and parallel motion becomes diffusive, leading to the compound sub-diffusion regime. Eventually, the diffusive regime in the perpendicular direction is restored when the particle displacement along the perpendicular direction is big enough to sample the fieldline complexity. A more detailed description of perpendicular transport scenarios is discussed in \citet{Shalchi2019}.


Figure~\ref{fig2} shows the parallel and perpendicular mean free paths, $\lambda_\parallel$ and $\lambda_\perp$, of all our simulations (circles) compared with the results of \citet{dundovic2020novel} (squares) and \citet{kuhlen2025simulations} (stars). For given values of $\zeta$, there is excellent agreement between our results and those of \citet{dundovic2020novel,kuhlen2025simulations}, despite the very different methods used. Our results are also in qualitative agreement with earlier work by \citep{husseinetal2015,giacinti2018reconciling}.
Our simulations achieve lower turbulence levels ($\zeta=0.3$) than previous literature, a rather challenging goal in terms of numerical computation.

 Figure~\ref{fig2} shows that $\lpa$ and $\lpe$ have different trends as a function of the particle energy. For $r_L>L_c$, $\lpa\propto(r_L/L_c)^2$, while $\lpe$ is flat. Parallel transport can be explained by quasi-linear theory  (QLT), or rather its second-order generalization that accounts for resonance broadening \citep{shalchi2005second}, whereas perpendicular transport arises from the random walk of field lines. For $r_L<L_c$, $\lpa$ follows a unique power law $\propto (r_L/L_c)^{1/3}$ for all the energies considered, while $\lpe$ has a double trend. In the range $10^{-2}<r_L/L_c<1$, $\lpe$ grows more rapidly than $\propto (r_L/L_c)^{1/3}$, thus differently from $\lpa$, while for $r_L/L_c<10^{-2}$, the same scaling as $\lpa$ is observed. This description applies to all the values of the turbulence amplitude considered here. 

\begin{figure}[h!]
 \centering
 \includegraphics[width=\hsize]{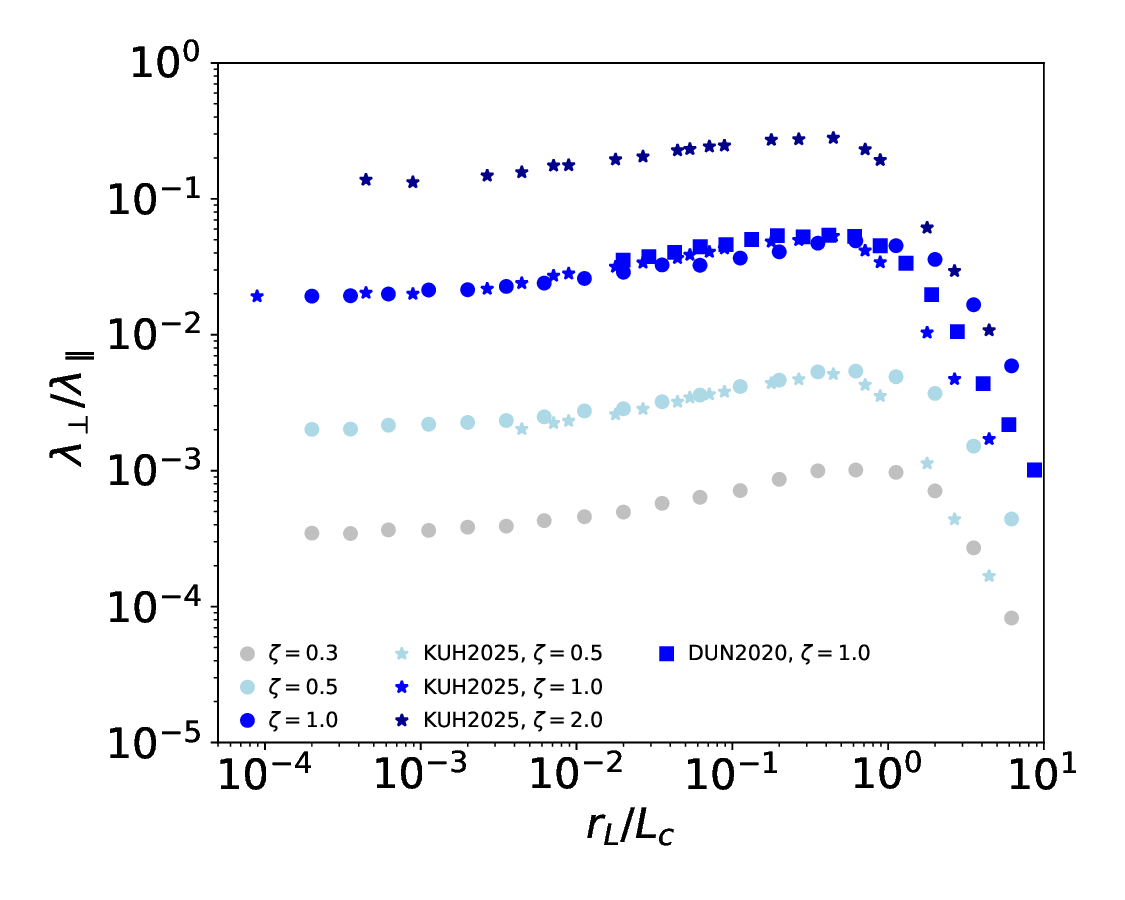}
   \caption{Perpendicular to this difference stems from how synthetic turbulence models populate the energy spectrum of turbulence levels: $\zeta=0.3$ (silver), $\zeta=0.5$ (light-blue), $\zeta=1.0$ (blue), $\zeta=2.0$ (darkblue). Different symbols correspond to this work (circles), \citet{kuhlen2025simulations} (KUH2025, stars), and \citet{dundovic2020novel} (DUN2020, squares).}
         \label{fig3}
 \end{figure}

The different trends of $\lpa$ and $\lpe$ as functions of energy become more evident when considering their ratio, as shown in Figure~\ref{fig3}, where we also compare our results with previous ones. 
While there is general agreement, we notice that for $r_L>L_c$, the $\lambda_\perp/\lambda_\parallel$ that we obtain is consistently larger than that obtained in previous calculations. This discrepancy arises from differences in the energy spectrum at large scales: different authors make different assumptions about the power spectrum at $k L_c<1$, and even within the same assumptions, the actual spectrum may depend on the specific method used to generate the low-$k$ modes and on the numerical resolution. In fact, none of the methods discussed here perfectly reproduces the trend $\propto k^4$ that would be required to fulfill the condition of homogeneity of the turbulence \cite[]{Batchelor1982}.


Overall, our simulation results corroborate previous findings on the different energy dependence of perpendicular and parallel diffusion coefficients, extending them to significantly smaller Larmor radii and amplitudes of turbulent magnetic fluctuations.

\section{A simple physical model}
\label{sec:simple}
As discussed in the previous section, numerical simulations of 3D isotropic turbulence show that parallel and perpendicular diffusion coefficients scale differently with particle energy for about two decades in particle gyroradii in the range $10^{-2}<r_L/L_c<1$. In particular, an increasing ratio $\lpe/\lpa$ is observed in that range as energy increases, confirming previous results in the same energy region \cite[]{DeMarco2007,husseinetal2015,giacinti2018reconciling,dundovic2020novel}.
Both the unified nonlinear theory (UNLT) and the nonlinear guiding center theory (NLGC) predict that the ratio $D_\perp/D_\parallel$ is constant at low energies and decreases at high energies (see \citealt{dundovic2020novel} and discussion below).
 

In the case of weak turbulence, $\delta B \ll B_0$, one can approximate the curvilinear element $ds$ as the displacement in the $z$ direction: $ds \approx dz$. If the guiding center of the particles follows the magnetic field lines, then one can write:
\begin{align}
\left\langle \Delta x^2\right\rangle
&\simeq \left\langle \Delta x^2\right\rangle_{\mathrm{FL}}=2\int_{0}^{s} ds'\, d_{\mathrm{FL}}(s')
\approx 2\int_{0}^{\sqrt{\left\langle \Delta z^2\right\rangle}} dz'\, d_{\mathrm{FL}}(z').
\end{align}
This implies that, from the definition of $d_\perp$ and $d_\parallel$:
\begin{align}
d_{\perp}(t)
&=\frac{1}{2}\frac{d}{dt}\left\langle \Delta x^2\right\rangle
=\frac{1}{2}\frac{d}{dt}\left(2\int_{0}^{\sqrt{\left\langle \Delta z^2\right\rangle}} dz'\, d_{\mathrm{FL}}(z')\right) \nonumber\\
&
=\frac{d_{\mathrm{FL}}\!\left(\sqrt{\left\langle \Delta z^2\right\rangle}\right)}{\sqrt{\left\langle \Delta z^2\right\rangle}}
\,d_{\parallel}(t)\label{dperpdpar}
\end{align}
which, as also reported by \citet{kuhlen2025models}, provides a direct mathematical relation between $\dperp$ and $\dpar$.
Notice that if we were in the purely diffusive regime of parallel diffusion and field line random walk (FLRW), then
$d_{\mathrm{FL}}\to \text{const}$ and $d_{\parallel}\to \text{const}$,
so that:
\begin{equation}
d_{\perp}\propto \frac{1}{\sqrt{\left\langle \Delta z^2\right\rangle}}
\sim t^{-1/2}.
\end{equation}
In this regime, particles would move subdiffusively in the perpendicular direction. Therefore, Equation (\ref{dperpdpar}) can only hold until the perpendicular displacement of the particles reaches $\sim L_c$, which is the perpendicular distance at which field lines decorrelate. When that distance is reached, the perpendicular diffusion coefficient $d_{\perp}$ saturates at
\begin{equation}
d_{\perp} (\tau_c)\sim \frac{L_c^2}{2\tau_c}.
\end{equation}
Considering that $\dperp(\tau_c)\simeq\Dperp^*$, the last expression implicitly defines $\tau_c$ as 
\begin{equation}\label{eq:tauc}
\tau_c \sim \frac{L_c^2}{2\Dperp^*}.    
\end{equation}
From $\tau_c$ and using $\dpar(\tau_c)\simeq\Dpar^*$, we can define the parallel displacement at $\tau_c$ as 
\begin{equation}\label{eq:sc}
s_c = s(\tau_c) \sim \sqrt{2 \Dpar^*\tau_c}    
\end{equation}
As we show below, the reasoning presented above leads to an energy-independent ratio of the perpendicular to the parallel diffusion coefficient when a ``standard'' ballistic-plus-diffusive behavior of magnetic field lines is taken into account, namely, the $d_{FL}$ is parametrized as
\begin{equation}\label{dflstandard}
         d_{FL}(s) =
        \begin{cases}
         \left(\frac{\delta B}{B}\right)^2s  & s \le L_c\\
         \left(\frac{\delta B}{B}\right)^2L_c  & s > L_c.\\
        \end{cases}
\end{equation}
Let us assume that $s_c>L_c$ and estimate the perpendicular diffusion coefficient (we will check {\it a posteriori} whether this assumption holds). 
By using Equation (\ref{eq:sc}), we can write Equation (\ref{dperpdpar}) at $t=\tau_c$ as
\begin{equation}
\Dperp^* = \frac{\left(\delta B/B\right)^2 L_c}{\sqrt{2\Dpar^*\tau_c}}\Dpar^*,
\end{equation}
and using Equation (\ref{eq:tauc}) we get
\begin{equation}
\frac{\Dperp^*}{\Dpar^*}
=\left(\frac{\delta B}{B}\right)^4
=\text{constant},
\label{eq:ratio}
\end{equation}
confirming previous results obtained by \citet{Shalchi2019}.
Notice that in this situation we obtain $s_c^2=2D_\parallel \tau_c=L_c^2/(\frac{\delta B}{B_0})^4\gg L_c^2$, if $\delta B/B_0\ll 1$, thereby confirming the initial assumption that $s_c>L_c$. Equation (\ref{eq:ratio}) shows that if the running diffusion coefficient of magnetic field lines can be approximated as in Equation (\ref{dflstandard}), then the ratio $D_\perp/D_\parallel$ is expected to be constant as a function of energy and proportional to $(\frac{\delta B}{B_0})^4$. 

\citet{kuhlen2025models} proposed that a growing ratio may result from a different parametrization of $d_{FL}$, with a subdiffusive phase for $s\gtrsim L_c$. 
The results strongly depend on the parametrization that is chosen, and the ones that seem to work the best are not in excellent agreement with the results of the numerical simulations of $d_{FL}$
(see Figure 2 in \citet{kuhlen2025models}). Moreover, although the results of our simulations are in excellent agreement with those of \citet{kuhlen2025simulations} for $\lpe/\lpa$, we do not observe a subdiffusive trend of $d_{FL}$ for all values of $\delta B/B_0$, that --in the theoretical treatment by \citet{kuhlen2025models}-- is the physical effect responsible for the energy-dependent $\lpe/\lpa$.

\begin{figure}[h!]
 \centering
 \includegraphics[width=\hsize]{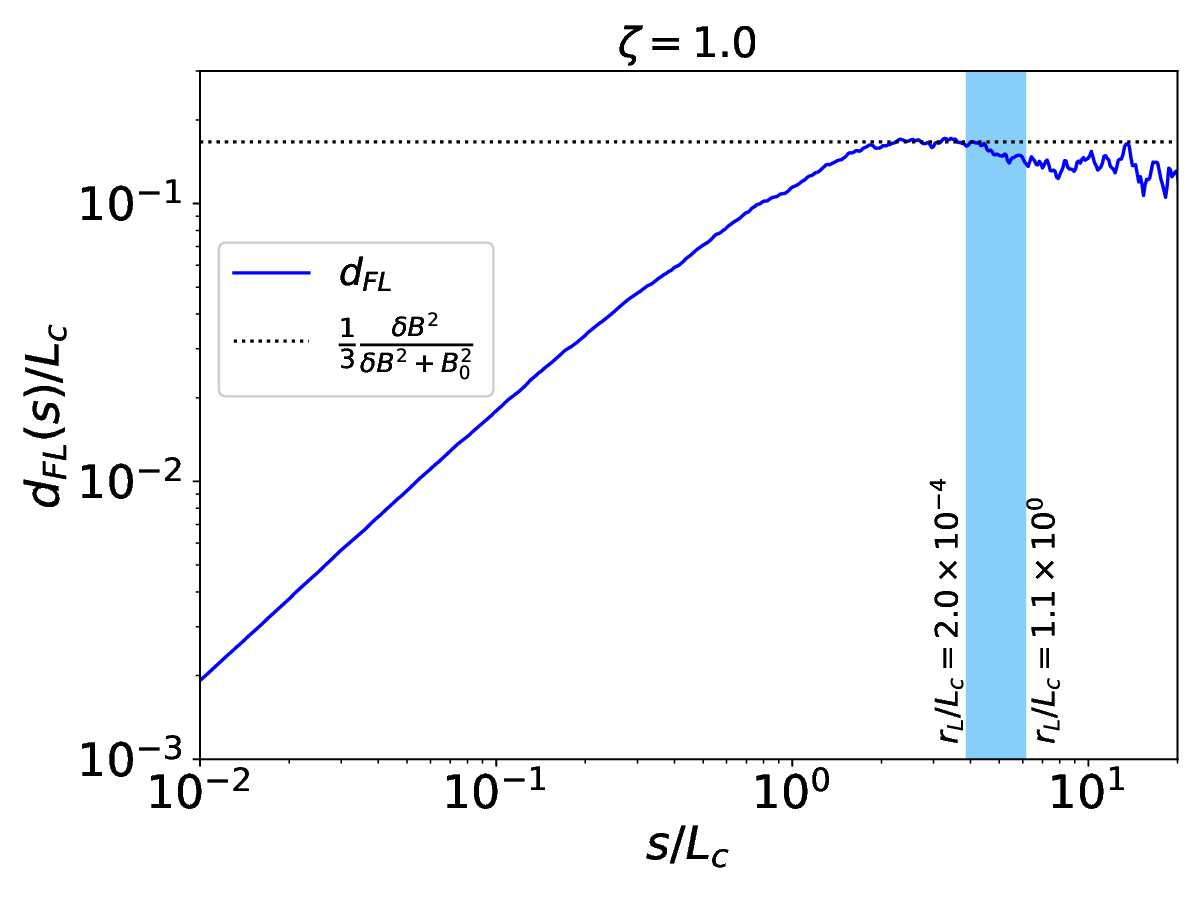}
   \caption{Field line running diffusion coefficient $d_{FL}(s)/L_c$ as a function of the field-line aligned coordinate $s/L_c$ for $\delta B/B_0=1$. The blue curve shows the measured $d_{FL}$, while the dotted horizontal line marks the theoretical saturation estimate, $\frac{1}{3}\frac{\delta B^2}{\delta B^2+B_0^2}$. The shaded vertical band indicates where $s_c$ falls for $2\times10^{-4}\le r_L/L_c \le 1.1$.}
         \label{fig4}
 \end{figure} 
For example, Figure~\ref{fig4} shows $d_{FL}$ for $\delta B/B_0=1$ computed from our synthetic turbulence model. It reaches saturation at $s \sim L_c$, while for longer distances it appears flat. If there is a subdiffusive phase, it seems to be very weak and insufficient to lead to a $D_\perp/D_\parallel$ increasing with energy. 
The blue shaded area in Figure~\ref{fig4} shows the range of values of $s_c$ as in Equation (\ref{eq:sc}) for $r_L/L_c$ ranging between $\sim 10^{-4}$ and 1. Again, one can easily appreciate that the range is so narrow that it is hard to think that it may give rise to a pronounced energy dependence of the $\lambda_\perp/\lambda_\parallel$ ratio. As a matter of fact, this result is easy to understand: perpendicular transport kicks in when the perpendicular displacement is of order $L_C$, but the corresponding parallel displacement is always of order $L_c/(\frac{\delta B}{B_0})^2$, which is basically independent of particle energy.


For these reasons, we believe that the subdiffusive phase of the magnetic field lines, if it has any effect on perpendicular transport, is not sufficient to account for the observed energy-dependent trend of $\lambda_\perp$. Another energy-dependent effect must be at play. In the next section, we propose a physical process that might provide a satisfactory alternative. 

\section{A NLGCT of perpendicular transport including stochastic drifts}
\label{sec:drifts}

A central idea in describing perpendicular transport is that, in the first approximation, particles follow magnetic field lines. In general, the position of the particle as a function of time can be written as 
\begin{equation}
  \bm x(t) = \R(t) + \vrho(t),
  \label{eq:x-R-rho}
\end{equation}
where \(\R\) is the guiding-center position and \(\vrho\) is the gyration vector. Since the gyromotion is perpendicular to the local magnetic field, in the limit of small magnetic perturbations $\delta B\ll B_0$, for a guide field along \(\zhat\),
\begin{equation}
  \vrho = \rho(\cos\phi\,\xhat + \sin\phi\,\yhat),
  \qquad
  \vrho\cdot\zhat = 0.
  \label{eq:rho-def}
\end{equation}
Below, we will use the Fourier notation for the perturbed magnetic field components:
\begin{equation}
  \delta B_i(\bm x) = \int \dd^3k\,\delta B_i(\kvec)\e^{\ii\kvec\cdot\bm x}.
  \label{eq:fourier-dB}
\end{equation}
If we write the wavevector as
\begin{equation}
  \kvec = \kvec_\perp + k_\parallel \zhat,
  \qquad
  \kvec_\perp = k_\perp(\cos\psi\,\xhat + \sin\psi\,\yhat),
  \label{eq:k-decomp}
\end{equation}
then
\begin{equation}
  \kvec\cdot\vrho = \kvec_\perp\cdot\vrho
  = k_\perp \rho \cos(\phi-\psi).
  \label{eq:kdotrho}
\end{equation}

To first order in \(\delta B/B_0\), the gyroaveraged guiding-center velocity can be written as
\begin{equation}
  \dot{\R} = \vpar\bhat + \bm v_{\nabla B} + \bm v_c,
  \label{eq:gc-velocity-decomp}
\end{equation}
where
\begin{equation}
  \bm v_{\nabla B}
  = \frac{\mu_m c}{qB}\,\bhat\times\nabla B,
  \qquad
  \bm v_c
  = \frac{\gamma m c\vpar^2}{qB}\,\bhat\times(\bhat\cdot\nabla)\bhat
  \label{eq:drifts}
\end{equation}
respectively describe the gradient and curvature drift motions, and we introduced the magnetic moment $\mu_m = \frac{p_\perp^2}{2\gamma m B_0}= \frac{\gamma m\vperp^2}{2B_0}$. 

Linearizing Equation (\ref{eq:gc-velocity-decomp}), the \(x\)-component becomes
\begin{equation}
v_x = v_x^{(x)} + v_x^{(z)} + v_x^{(y)},
\label{eq:vx-decomp}
\end{equation}
with
\begin{align}
  v_x^{(x)}
  &= \frac{\vpar}{B_0}\delta B_x(\R+\vrho),
  \label{eq:vx-17}\\
  v_x^{(z)}
  &= -\frac{\mu_m c}{qB_0}\,\partial_y\delta B_z(\R+\vrho),
  \label{eq:vx-18}\\
  v_x^{(y)}
  &= -\frac{\gamma m c\vpar^2}{qB_0^2}\,\partial_z\delta B_y(\R+\vrho),
  \label{eq:vx-19}
\end{align}

where for simplicity we defined the different terms $v_x^{(j)}$ as they depend on different components of magnetic field perturbations $\delta B_j$, with $j=x,y,z$. 
The $x-$component of the guiding center velocity $\widetilde v_x$ (relevant for the calculation of the perpendicular diffusion coefficient) is obtained by averaging $v_x$ over a gyroperiod. Using its Fourier representation, $\widetilde v_x$ can be written as:
$$
  \widetilde v_x(t)
  = \int \dd^3k\,\e^{\ii\kvec\cdot\R(t)}J_0(k_\perp\rho) \times
$$
\begin{equation}
 \times \left[
    \frac{\vpar(t)}{B_0}\delta B_x(\kvec)
    - \frac{\ii\rho\vperp(t)}{2B_0}k_y\delta B_z(\kvec)
    - \frac{\ii\rho\vpar^2(t)}{B_0\vperp(t)}k_z\delta B_y(\kvec)
  \right].
  \label{eq:vx-fourier-rho}
\end{equation}
where we used that
\begin{equation}
  \gyroavg{\e^{\ii\kvec\cdot(\R+\vrho)}}
  = \e^{\ii\kvec\cdot\R}\frac{1}{2\pi}\int_0^{2\pi}\dd\theta\,
    \e^{\ii k_\perp\rho\cos\theta} = \e^{\ii\kvec\cdot\R}J_0(k_\perp\rho),
\end{equation}
and $\gyroavg{\ldots}$ denotes the average over a gyroperiod, $\rho = \frac{\vperp}{\Omega} = \frac{p_\perp c}{qB_0}$, and $J_0$ is a Bessel function.  

At this point, we can define the perpendicular diffusion coefficient as 
\begin{equation}
    D_{xx} = \int_0^\infty dt \,\langle \tilde v_x (t) \tilde v_x (0)\rangle = \int_0^\infty dt \,C_{xx}(t),
    \label{eq:FullDxx}
\end{equation}
where $\langle\ldots\rangle$ means averaging over the ensemble particles. Using Equation \eqref{eq:vx-fourier-rho} and after some math, one arrives at:
\begin{equation}
  C_{xx}(t)
  = \int \dd^3k\,J_0^2(k_\perp\rho)
  \avg{\e^{\ii\kvec\cdot\Delta\R(t)}S_{xx}(\kvec,t)},
  \label{eq:Cxx-full}
\end{equation}
with $\Delta\R(t)=\R(t)-\R(0)$ and

\begin{align}
  &S_{xx}(\kvec,t)
  = \frac{\vpar(t)\vpar(0)}{B_0^2}P_{xx}(\kvec)
   + \frac{\rho^2\vperp(t)\vperp(0)}{4B_0^2}k_y^2P_{zz}(\kvec)
  \notag\\
  &\quad
   + \frac{\rho^2\vpar^2(t)\vpar^2(0)}{B_0^2\vperp(t)\vperp(0)}k_z^2P_{yy}(\kvec)
  \notag\\
  &\quad
   + \frac{\ii\rho}{2B_0^2}\vpar(t)\vperp(0)k_yP_{xz}(\kvec)
   - \frac{\ii\rho}{2B_0^2}\vperp(t)\vpar(0)k_yP_{zx}(\kvec)
  \notag\\
  &\quad
   + \frac{\ii\rho}{B_0^2}\frac{\vpar(t)\vpar^2(0)}{\vperp(0)}k_zP_{xy}(\kvec)
   - \frac{\ii\rho}{B_0^2}\frac{\vpar^2(t)\vpar(0)}{\vperp(t)}k_zP_{yx}(\kvec)
  \notag\\
  &\quad
   - \frac{\rho^2}{2B_0^2}\frac{\vperp(t)\vpar^2(0)}{\vperp(0)}k_yk_zP_{zy}(\kvec)
   - \frac{\rho^2}{2B_0^2}\frac{\vpar^2(t)\vperp(0)}{\vperp(t)}k_yk_zP_{yz}(\kvec),
  \label{eq:Sxx-full}
\end{align}
To get Equations (\ref{eq:Cxx-full})-(\ref{eq:Sxx-full}), we have factorized the magnetic field terms in the correlator and used the condition of homogeneous turbulence $\avg{\delta B_i(\kvec)\delta B_j^*(\kvec')}= P_{ij}(\kvec)\delta(\kvec-\kvec')$. For isotropic turbulence, the $P_{ij}$ tensor can be written as:
\begin{equation}
  P_{ij}(\kvec) = P(k)\left(\delta_{ij}-\frac{k_i k_j}{k^2}\right),
  \label{eq:isotropic-Pij}
\end{equation}
where $P(k)$ is related to the power spectrum $\tilde P(k)$ of turbulence by $\delta B^2/2 = \int dk 4\pi k^2 P(k) = (1/2) \int dk \tilde P(k)$, namely $\tilde P(k)/2=4\pi k^2 P(k)$. Eliminating all terms that are odd in one of the components of $k$ and do not contribute to the integral, one has
\begin{align}
  S_{xx}(\kvec,t)
  &= P(k) \left\{\frac{\vpar(t)\vpar(0)}{B_0^2}\left(1-\frac{k_x^2}{k^2}\right)
  \notag \right.\\
  &\left.\quad
   + \frac{\rho^2\vperp(t)\vperp(0)}{4B_0^2}k_y^2\frac{k_\perp^2}{k^2}
  \notag\right.\\
  &\left. \quad
   + \frac{\rho^2\vpar^2(t)\vpar^2(0)}{B_0^2\vperp(t)\vperp(0)}k_z^2\left(1-\frac{k_y^2}{k^2}\right)
  \notag\right.\\
  &\left.\quad
   + \frac{\rho^2}{2B_0^2}
   \left[
     \frac{\vperp(t)\vpar^2(0)}{\vperp(0)}
     + \frac{\vpar^2(t)\vperp(0)}{\vperp(t)}
   \right]\frac{k_y^2k_z^2}{k^2} \right\}.
  \label{eq:Sxx-surv}
\end{align}
In this expression, the last term involves mixed correlations between the parallel and perpendicular velocities, which are usually negligible; in fact, we checked that, by treating these terms in several ways, the last term can always be neglected. Hence, in the following, this term will be ignored. The first term describes the standard perpendicular transport due to particles following the field lines. The second and third terms describe gradient and curvature drifts, respectively. In these last two terms, we will make the simplifying assumption that $v_\parallel^2(t)=v_\parallel^2(0)=v^2/3$ and $v_\perp(t)v_\perp(0)\approx \avg{v_\perp^2}\approx (2v^2/3)$.

\subsection{Standard $D_\perp$}
Using Equation (\ref{eq:FullDxx}), Equation (\ref{eq:Cxx-full}) and the first term in Equation (\ref{eq:Sxx-surv}), the standard perpendicular diffusion coefficient in the $x$-direction can be written as 
\begin{equation}
D_{xx}=\int_0^\infty dt \,\int \dd^3k\,J_0^2(k_\perp\rho) P_{xx}(\kbold)
  \frac{\avg{\e^{\ii\kvec\cdot\Delta\R(t)}\vpar(t)\vpar(0)}}{B_0^2}.
  \label{eq:Dxx}
\end{equation}

The term in $\avg{...}$ can be calculated in a couple of different ways, and, rather surprisingly, they yield sensibly different results. One way is to adopt the strategy proposed by \cite{matthaeus2003nonlinear} and split the 
correlator into the product of two 
correlators. 
Assuming an exponential decorrelation in the parallel direction, $\langle v_\parallel(t) v_\parallel(0)\rangle=(v^2/3)e^{-vt/\lambda_\parallel}$, where $\lambda_\parallel$ is the pathlength for parallel diffusion. If diffusion is assumed in all three spatial directions, then it is easy to show that 
\begin{equation}
    \avg{\e^{\ii\kvec\cdot\Delta\R(t)}} = e^{-D_\parallel k_\parallel^2 t - D_\perp k_\perp^2 t},
    \label{eq:DiffClosure}
\end{equation}
and replacing these results in Equation (\ref{eq:Dxx}):
\begin{equation}
    D_{xx}=\frac{a^2 v^2}{3 B_0^2}\int d^3 k \frac{P_{xx}(\kbold)J_0^2(k_\perp\rho)}{D_\parallel k_\parallel^2+D_\perp k_\perp^2 +\frac{v}{\lambda_\parallel}}.
\end{equation}
Following \cite{matthaeus2003nonlinear,ShalchiBook2009}, we introduced a numerical factor $a^2$ of order unity, which should account for mild violations of the assumption that particles follow magnetic field lines. We will retain this factor here, stressing, however, that NLGCT has never been applied to isotropic turbulence, except \cite{dundovic2020novel}, where it was found that it does not describe the simulation results well, a finding that we confirm here and comment on below.

\cite{shalchi2010aunified} proposed the so-called unified non-linear (UNL) approach to perpendicular diffusion in which the correlator $\avg{\e^{\ii\kvec\cdot\Delta\R(t)}\vpar(t)\vpar(0)}$ is calculated using the Vlasov equation and leads to:
\begin{equation}
D_{xx}^{\rm UNL}=\frac{a^2 v^2}{3 B_0^2}\int d^3 k \frac{P_{xx}(\kbold)J_0^2(k_\perp\rho)}{\frac{v}{\lambda_\parallel}+\frac{4}{3}D_\perp k_\perp^2 +\frac{(v k_\parallel)^2}{3 D_\perp k_\perp^2}}.    
\end{equation}
As for NLGCT, UNLT does not align well with the simulation results for isotropic turbulence \citep{dundovic2020novel}.

We checked that neither of the two expressions for $D_{xx}$ provides an accurate description of our numerical results over the entire energy range. Both UNLT and NLGCT reproduce the numerical results well at the lowest energies. However, the UNLT prediction starts to deviate from the simulations only at higher energies than the standard NLGCT prediction. Therefore, UNLT remains accurate over a broader energy range than NLGCT.
Moreover, it can be shown that for isotropic turbulence, NLGCT leads to a scaling of $D_\perp/D_\parallel$ with $\zeta$ that is inconsistent with simulation results at low energies. Hence, in the following, we will use the UNLT theory to describe standard perpendicular diffusion. Clearly, since turbulence is isotropic, we can write $D_\perp=(D_{xx}+D_{yy})/2$. Both the results of the UNLT approach and of NLGCT are expressed as nonlinear integral equations, which are solved iteratively for each value of the particle energy. More explicitly, the standard diffusion, assuming that particles follow magnetic field lines, is:
\begin{equation}
D_\perp^{\rm st}=\frac{a^2 v^2}{6 B_0^2}\int d^3 k \frac{P(k)\left(1+\frac{k_\parallel^2}{k^2}\right)}{\frac{v}{\lambda_\parallel}+\frac{4}{3}D_\perp k_\perp^2 +\frac{(v k_\parallel)^2}{3 D_\perp k_\perp^2}}J_0^2(k_\perp\rho).    
\end{equation}

\subsection{Perpendicular transport due to stochastic drifts}

Perpendicular diffusion due to gradient and curvature drifts is due to the second and third terms in Equation (\ref{eq:Sxx-surv}). Summing the two components and using the symmetry $D_{xx}=D_{yy}$, we get:
\begin{align}
D_\perp^{\rm drift}=&\frac{\rho^2 v^2}{12 B_0^2}\int d^3 k \, P(k) k^2 \left(1+2\frac{k_\parallel^4}{k^4}-\frac{k_\parallel^2}{k^2}\right)\times\nonumber\\
&\times J_0^2(k_\perp\rho)\int_{0}^\infty dt \avg{\e^{\ii\kvec\cdot\Delta\R(t)}}.  
\label{eq:Ddrift}
\end{align}
At this point, we face the same issue as above concerning the closure to adopt in the calculation of $\avg{\e^{\ii\kvec\cdot\Delta\R(t)}}$. We consider two alternatives that yield slightly different descriptions of the role of drifts: the first is the diffusive closure, Equation (\ref{eq:DiffClosure}), that leads to
\begin{equation}
    G_{\rm diff}(k_\perp,k_\parallel) = \int_0^\infty dt \avg{\e^{\ii\kvec\cdot\Delta\R(t)}} =\frac{1}{D_\parallel k_\parallel^2+D_\perp k_\perp^2}.
\end{equation}
This closure assumes that decorrelation in the parallel and perpendicular directions occurs on their respective diffusive time scales. On the other hand, from simulations we observe that, especially for low values of $\delta B/B$, when the particles move a distance $\sim L_c$ in the perpendicular direction, they have moved by less than a path length in the direction parallel to $B_0$. This transition between quasi-ballistic and diffusive behavior is best captured by the Telegraph equation  \citep{litvinenkoandschlicheiser2013,tautzandlerche2016,shalchi2024} (see Appendix \ref{app:telegraph}). In this case, we find:
\begin{equation}
    \avg{\e^{\ii\kvec\cdot\Delta\R(t)}} = e^{- D_\perp k_\perp^2 t} T(k_\parallel,t),
    \label{eq:TeleClosure}
\end{equation}
with 
\begin{equation}
    T(k_\parallel,t)=e^{-t/2\tau_\parallel}\left[ \cosh(\omega t)+\frac{1}{2\tau_\parallel\omega}\sinh(\omega t)\right],
\end{equation}
where $\omega^2=\frac{1}{4\tau_\parallel^2}-\frac{v^2 k_\parallel^2}{3}$. After some algebra, one can show that 
\begin{equation}
    G_{\rm tel}(k_\perp,k_\parallel) = \int_0^\infty dt \avg{\e^{\ii\kvec\cdot\Delta\R(t)}} =\frac{1+\tau_\parallel D_\perp k_\perp^2}{D_\perp k_\perp^2+\tau_\parallel D_\perp^2 k_\perp^4 +\frac{1}{3}v^2 k_\parallel^2}.
\end{equation}
The perpendicular diffusion coefficient due to the combined effect of gradient and curvature drifts can be obtained by replacing either $G_{\rm diff}$ or $G_{\rm tel}$ into Equation (\ref{eq:Ddrift}).  



\section{Theory versus simulations}
\label{sec:TS}

\begin{figure}[h!]
\centering
\includegraphics[width=\hsize]{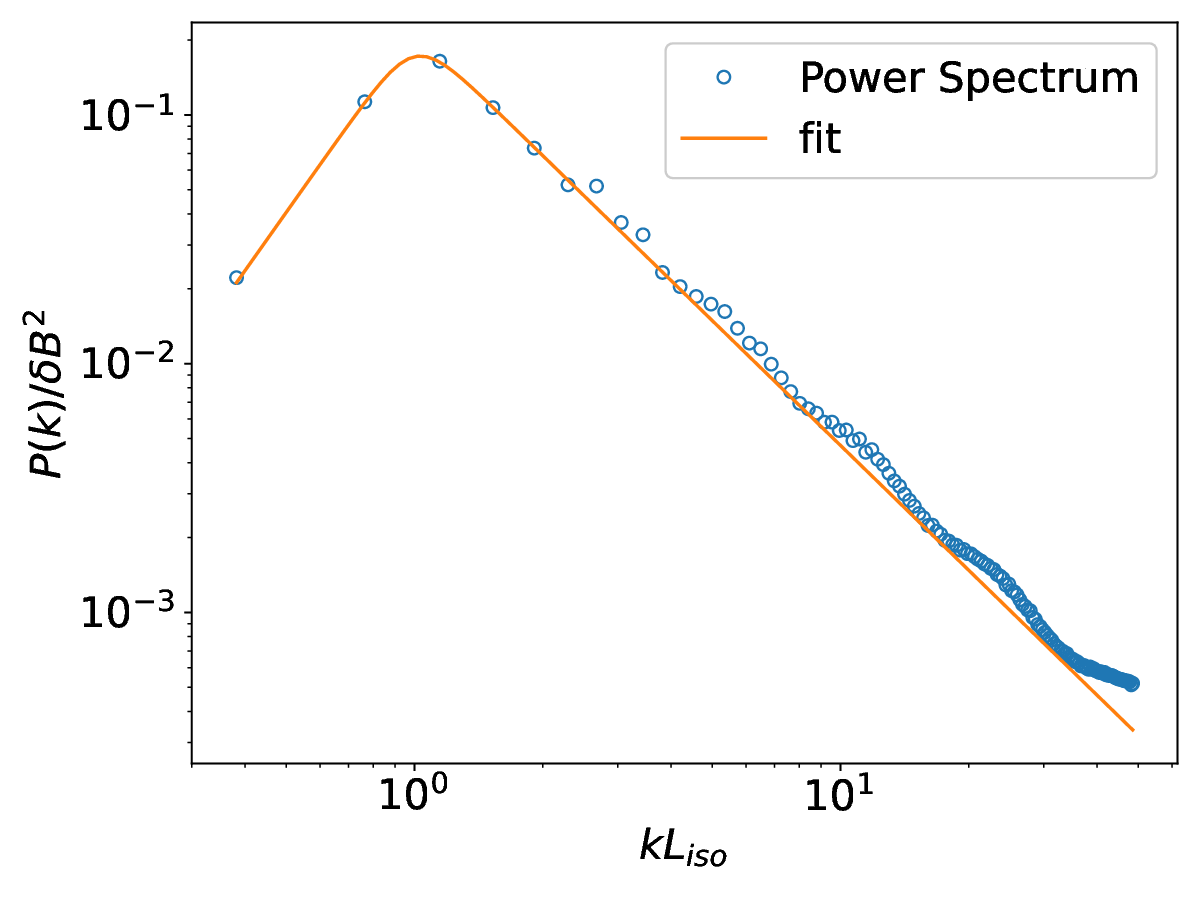}
\caption{Power spectrum from the synthetic turbulence model at low $k$ and fit. The power spectrum is computed from a magnetic field generated by the synthetic turbulence model on a cubic 3D grid of $256^3$ equally spaced points. To produce this plot, a reduced spectral length of 2 orders of magnitude is used to avoid spectral aliasing. The oscillation observed in the spectra at high $k$ values is smoothed out when the spectra are computed on different grids and then averaged (see e.g., Figure 6 in \citet{malara2016fast}). The actual spectrum used in the test-particle simulations spans 7 orders of magnitude and overlaps with the spectrum reported here, extending to higher $k$ values and following a $-5/3$ spectral slope.}
\label{fig:spectrum}
\end{figure}

The full equation for perpendicular transport can now be written as:
\begin{align}
D_\perp&=\frac{a^2 v^2}{6 B_0^2}\int d^3 k \frac{J_0^2(k_\perp\rho) P(k)\left(1+\frac{k_\parallel^2}{k^2}\right)}{\frac{v}{\lambda_\parallel}+\frac{4}{3}D_\perp k_\perp^2 +\frac{(v k_\parallel)^2}{3 D_\perp k_\perp^2}} +\nonumber\\
&\frac{\rho^2 v^2}{12 B_0^2}\int d^3 k \, P(k) k^2 \left(1+2\frac{k_\parallel^4}{k^4}-\frac{k_\parallel^2}{k^2}\right)J_0^2(k_\perp\rho)G(k_\perp,k_\parallel),
\label{eq:golden}
\end{align}
where $G(k_\perp,k_\parallel)$ is $G_{\rm diff}$ or $G_{\rm tel}$ depending on the closure, and $\rho$ should be intended as related to $r_L$ through $\rho=\sqrt{2/3}r_L$. With the normalization of the power spectrum discussed above, we have $P(k)=\tilde P(k)/8 \pi k^2$. Our synthetic turbulence model produces a $-5/3$ spectrum for scales smaller than the bendover scale. At larger scales, the spectrum is determined by the largest wavelet in the model. We fit the power spectrum in our simulations as
\begin{equation}
    \tilde P(k) = C L_{iso} \delta B^2 F(x),
\end{equation}
where $L_{iso}$ is the bend-over scale for isotropic turbulence and $x=k L_{iso}$ is the dimensionless scale in units of $L_{iso}$. To guarantee the homogeneity of turbulence, one should make sure that $F(x)\propto x^\alpha$, with $\alpha>4$ at $x\ll 1$ \cite[]{Batchelor1982}. On the other hand, numerical techniques do not always allow for obtaining this outcome, either because reproducing the region where this scaling would be achieved would require shifting the spectrum to lower $k$s, thus sacrificing the inertial range extension (as in \citealt{dundovic2020novel}), or because the method of generating turbulence forces a different $k$ dependence, as is our wavelet-based model. In Figure~\ref{fig:spectrum} we show the function $\tilde P(k)/\delta B^2$ (dots) as generated following the method discussed in Section \ref{sec:simula}, together with an analytic fit in the form:
\begin{equation}
\tilde P(k) = C L_{iso} \delta B^2 \frac{x^\alpha}{\left(1+x^\beta\right)^{\frac{\alpha+5/3}{\beta}}},
\end{equation}
with the normalization constant
\begin{equation}
    C=\beta \frac{\Gamma\left(\frac{\alpha+5/3}{\beta}\right)}{\Gamma\left(\frac{\alpha+1}{\beta}\right)\Gamma\left(\frac{2}{3\beta}\right)}.
\end{equation}

The parameters are fit to the simulation results and found to be $\alpha=2.42$ and $\beta=11.86$, while the numerical factor 5/3 is specialized here to the case of Kolmogorov turbulence at high $k$. The coherence scale $L_c$ calculated using the standard expression is \cite[]{dundovic2020novel}:
\begin{equation}
    L_c = \frac{1}{\delta B^2} \int_0^\infty\int_0^\infty dk \tilde P(k) \frac{\sin (kr)}{kr} = \frac{\pi}{2}\frac{\Gamma\left(\frac{\alpha}{\beta}\right)\Gamma\left(\frac{5}{3\beta}\right)}{\Gamma\left(\frac{\alpha+1}{\beta}\right)\Gamma\left(\frac{2}{3\beta}\right)}\, L_{iso}.
\end{equation}
The parameter $\beta$ regulates the sharpness of the transition around $L_{iso}$. The fact that $\beta$ is rather large implies that the transition is sharp, hence $L_c$ is close to $L_{iso}$, $L_c\approx 0.87 L_{iso}$. 

A few comments on Equation (\ref{eq:golden}) are in order: 1) the first term, that accounts for particles following magnetic field lines (standard contribution to perpendicular transport) is similar but not identical to the usual contribution in NLGC or UNL theory. It also contains the integral of the Bessel function $J_0$ related to the average magnetic field over particle orbit, so far neglected in the literature, and performed assuming that $\delta B_x$ is constant across the trajectory of the particle in the perpendicular plane. This correction is negligible for low-energy particles ($r_L/L_c\ll 1$). In contrast, at higher energies, it provides a small correction to UNL theory predictions, accounting for the fact that particles cannot perfectly follow magnetic field lines. In any case, none of the calculations described here should be used for $r_L/L_c\gtrsim 1$. 

2) The coefficient $a^2<1$ in the first term of Equation (\ref{eq:golden}) is the same as in the standard NLGC and UNL theory, and is introduced to mimic ad hoc the fact that particles do not exactly follow magnetic field lines. Interestingly, while in standard NLGC and UNL theory it is argued that $a^2=1/3$ is needed to reproduce simulation results, our calculations typically yield $a^2\sim 1/2-2/3$, depending on the value of $\zeta$. 

3) One can calculate the asymptotic behavior of the two terms in Equation (\ref{eq:golden}), in the limit of $r_L/L_c\ll 1$. In this limit, as we discuss below, the standard term dominates the perpendicular transport, with particles following field lines. However, we are interested in characterizing how the drift terms compete with standard diffusion. In the low energy limit, we find that $D_\perp/D_\parallel\propto \zeta^4$ for the standard perpendicular transport, independent of energy, as previously found by \citet{Shalchietal2015,Shalchi2019}. Hence, at low enough energies, the energy dependence of $D_\perp$ is expected to be the same as $D_\parallel$ and its normalization suppressed by roughly $\zeta^4$ with respect to the parallel diffusion coefficient. Both predictions are confirmed by our simulation results (see below). For the drift terms, the asymptotic behavior depends on the closure: for a diffusive closure, we find that $D_\perp^{\rm drift}\propto \zeta^2 \rho^2/D_\parallel$, and recalling that $D_\parallel\propto \rho^{1/3}/\zeta^2$ in the relevant limit, one obtains $D_\perp^{\rm drift}\propto \zeta^4 \rho^{5/3}$. For the telegraph closure, if we define the standard ratio as $R_{\rm ST}\propto\zeta^4$, one can write $D_\perp^{\rm drift}\propto \zeta^2\rho^2(R_{\rm ST} D_\parallel)^{-1/2}\propto\zeta \rho^{11/6}$.

Aside from asymptotic behavior, we solve Equation (\ref{eq:golden}) numerically using an iterative method that starts with a guess for $D_\perp$ and eventually converges to the solution, regardless of the initial guess. Typically, the method converges within $\lesssim 10$ iterations for each energy bin. As with standard NLGCT, the generalized approach, including stochastic drifts, requires knowledge of the parallel diffusion coefficient for the calculation. In our calculations, we use an analytic fit to $D_\parallel$ as a function of energy as obtained from the same simulations discussed in Section \ref{sec:simula}, for selected values of $\zeta=0.3,\;0.5,\;1$ (see Appendix \ref{app:A1}). 

\begin{figure*}[h!]
     \sidecaption
     \includegraphics[width=\textwidth]{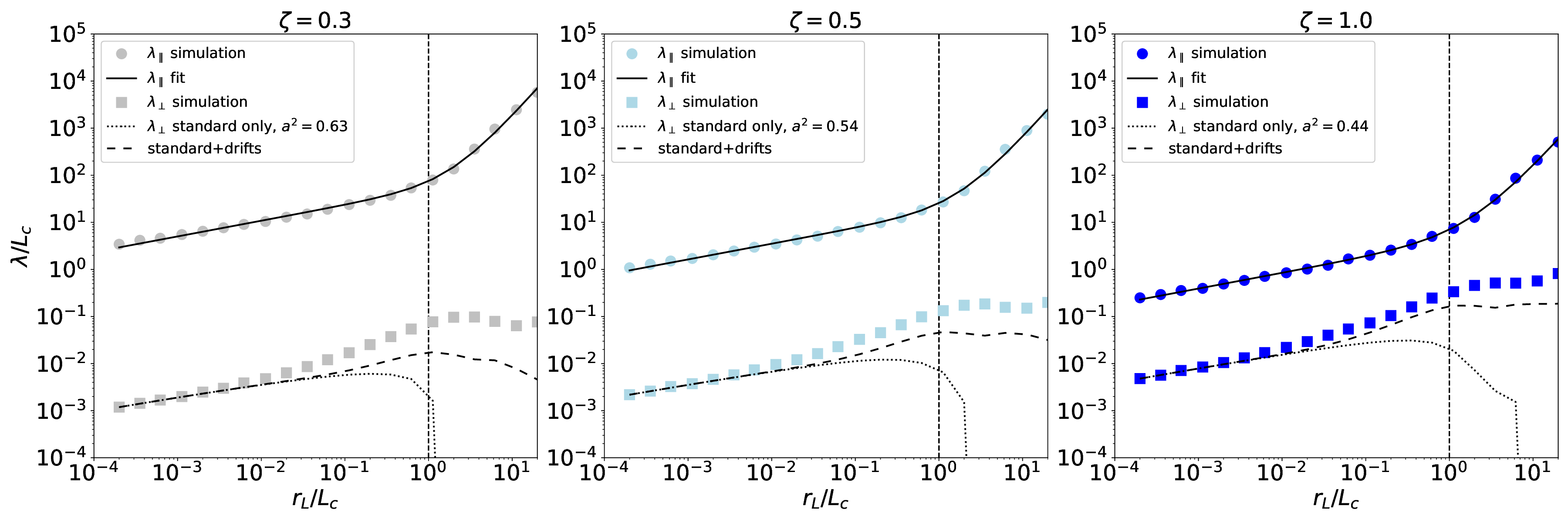}
   \caption{Parallel and perpendicular mean free path from simulations (circles and squares); fit of the parallel mean free path (solid line), perpendicular mean free path from standard field line diffusion (loosely dashed line), and from standard diffusion plus drifts (dashed line), using a diffusive closure. The three columns represent the cases $\delta B/B=0.3,0.5,1.0$.}
         \label{fig:fig6}
\end{figure*}

\begin{figure*}[h!]
     \sidecaption
     \includegraphics[width=\textwidth]{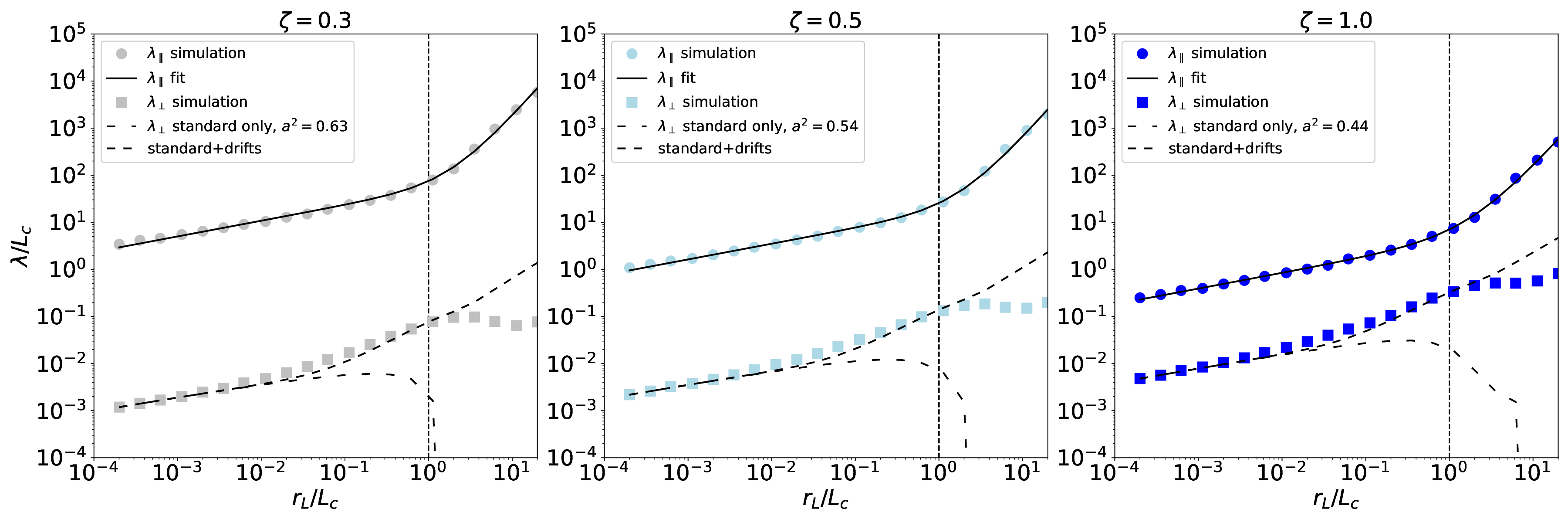}
   \caption{Parallel and perpendicular mean free path from simulations (circles and squares); fit of the parallel mean free path (solid line), perpendicular mean free path from standard field line diffusion (loosely dashed line), and from standard diffusion plus drifts (dashed line), using the telegraph closure. The three columns represent the cases $\zeta =0.3,0.5,1.0$.}
         \label{fig:fig7}
\end{figure*}

The results of the numerical calculations are shown in Figure~\ref{fig:fig6} for a diffusive closure and in Figure~\ref{fig:fig7} for the closure based on the telegraph equation. 

The diffusive closure seems to lead to too fast decorrelation in the parallel direction, which results in the drift terms being too small, especially for low values of $\zeta$. Our simulations confirm this suspicion: typically, perpendicular decorrelation occurs when the particles have moved by $\sim L_c$ in the perpendicular direction. Because of the random walk of field lines, this requires that in the parallel direction the particles have moved by $\sim L_c/\zeta^2$. Assuming diffusion on such time scales implies requiring that $L_c/\zeta^2\gg \lambda_\parallel$, but for $r_L/L_c\gtrsim 10^{-2}$ this condition is not fulfilled. This can be checked directly using the fit to our parallel diffusion coefficient in Appendix \ref{app:A1} (notice that $D_0$ scales approximately as $\propto \zeta^{-2}$, hence the value of $r_L/L_c$ where the condition above fails is the same for all values of $\zeta$). 

This regime, in between ballistic and diffusive, is exactly the one in which the telegraph equation provides a better description than the diffusion equation. The pathlength shown in Figure~\ref{fig:fig7} confirms that $\lambda_\perp$ calculated in this way provides a better description of the results of simulations. In particular, it leads to a stronger energy dependence of $\lambda_\perp$ for $r_L/L_c\gtrsim 10^{-2}$ due to the larger contribution of gradient and curvature drifts to perpendicular transport in this regime. 
 
\section{Conclusions}
\label{sec:conclude}

We ran numerical simulations of particle transport in synthetic homogeneous isotropic turbulence in the presence of a large-scale magnetic field with strength $B_0$, using the wavelet method put forward by \cite{malara2016fast}, which allowed us to extend the dynamical range by about two orders of magnitude with respect to previous simulations reconstructing the field on a grid \cite[]{DeMarco2007,dundovic2020novel,giacinti2018reconciling}. The wavelet method is competitive with the nested box technique presented by \cite{kuhlen2025models}. We used our simulations to compute the parallel and perpendicular diffusion coefficients as a function of particle rigidity, using $r_L/L_c$ as a proxy. We confirm the results of previous work for parallel diffusion: for a Kolmogorov spectrum of the turbulent magnetic field, we find that $D_\parallel \propto E^{1/3}$ with a normalization that can be estimated reasonably well using QLT (or rather its second-order generalization \citet{shalchi2005second}). At $r_L/L_c\gg 1$, we confirm the universal behavior $D_\parallel \propto E^2$, independent of the spectrum of turbulence. This behavior, first derived for slab turbulence \citep{jokipii1966cosmic}, was shown to be present even in the extreme case of isotropic turbulence in the absence of $B_0$ by \citet{subedi2017charged}. 

Our results on the perpendicular diffusion revealed an interesting result, already shown by the numerical simulations of \cite{kuhlen2025simulations,kuhlen2025models}. However, our results extend over a slightly larger dynamical range and lower levels of turbulence: on one hand, we confirm the finding of \cite{DeMarco2007} that $D_\perp/D_\parallel$ is not flat for $r_L/L_c\gtrsim 10^{-2}$, a result later found also by \citet{husseinetal2015,giacinti2018reconciling,dundovic2020novel}. But we find that the ratio flattens for $r_L/L_c\lesssim 10^{-2}$, almost independent of the value of $\delta B/B_0$, as previously demonstrated by \cite{kuhlen2025models}.

\cite{kuhlen2025models} suggested that this change of regime may be attributed to subdiffusion of magnetic field lines. However, we noted that the odd behavior also occurs for $\delta B/B_0=1$, where no clear signature of subdiffusion in the magnetic field lines can be identified in our simulations. Moreover, we show that the distance $s$ that the particles cover in the parallel direction before experiencing a perpendicular displacement of $\sim L_c$ is almost independent of energy and equal to $L_c/\zeta^2$. Hence, it is difficult to imagine why particles with different energies should experience different field-line random walks. In other words, it is natural to expect that $D_\perp/D_\parallel$ is constant for $r_L/L_c\lesssim 1$.

We explored alternative explanations for the energy dependence of $D_\perp$, using the NLGCT framework proposed by \cite{matthaeus1990evidence} and subsequent modifications \cite[]{ShalchiBook2009}.

The main novelty of the analytic approach discussed here is in the inclusion of the effects of stochastic drifts in the motion of the guiding center of the particles: the random gradient and curvature of the local magnetic field that a particle experiences in one gyration cause small deviations of the trajectory of the guiding center from perfectly following magnetic field lines. This effect contributes to the perpendicular transport, which depends more strongly on energy than the standard term. This results in drifts competing with the random walk of magnetic field lines for $10^{-2}\lesssim r_L/L_c\lesssim 1$. 

The main uncertainty in the calculations, including stochastic drifts, is the same as in the standard NLGCT: the closure. For instance a diffusive closure for $\avg{\e^{\ii\kvec\cdot\Delta\R(t)}}$ was used in NLGCT, while \cite{shalchi2010aunified} proposed to calculate $\avg{v_z v_z\e^{\ii\kvec\cdot\Delta\R(t)}}$ using Vlasov equation. These two prescriptions return substantially different results for standard perpendicular transport. 

The drift contribution contains a similar $\avg{\e^{\ii\kvec\cdot\Delta\R(t)}}$, and we computed it in two ways, which we refer to as the diffusive and the telegraph closure. 
The latter, never used in previous literature in this context, describes propagation using the telegraph equation rather than the standard diffusion equation and allows us to describe better situations in which decorrelation due to drifts occurs when the particles are not in the fully diffusive regime. 

We find that both methods yield a $D_\perp$ that is in better agreement with the simulation results, but the telegraph closure is quantitatively more accurate. 

Although more investigation of this intriguing problem is needed to explore additional details, we think that the long-sought-after explanation of the anomalous energy dependence of $D_\perp$ may be due mainly to the effect of random drifts. However, we cannot exclude that the subdiffusion of magnetic field lines, discussed by \cite{kuhlen2025models}, may also play a role, especially for low values of $\zeta$. Moreover, we wish to stress that the main contribution of the drifts is due to the $z$-component of the turbulent field, and this component is especially important in cases of magneto-sonic turbulence recently explored by \cite{Bouchet2026}. Hence, we are planning to extend the calculations presented here to the case of fast magnetosonic turbulence, where stochastic drifts might play an even more pronounced role. 

\begin{acknowledgements}
We acknowledge EuroHPC JU for awarding this project access to MeluXina CPU at LuxProvide (LU) through the grant EHPC-REG-2022R03-217.
FP and FM acknowledge the project ‘Data-based predictions of solar energetic particle arrival to the Earth: ensuring space data and technology integrity from hazardous solar activity events’
(CUP H53D23011020001) ‘Finanziato dall’Unione europea – Next Generation EU’ PIANO NAZIONALE DI RIPRESA E RESILIENZA (PNRR) Missione 4 “Istruzione e Ricerca” - Componente C2 Investimento 1.1, ‘Fondo per il Programma Nazionale di Ricerca e Progetti di Rilevante Interesse Nazionale (PRIN)’ Settore PE09. The research of PB and CE was partially funded under PRIN-MUR 2022TJW4EJ ``Unveiling the footprints of the cosmic ray journey through the Galaxy and beyond'' and under MUR National Innovation Ecosystem grant ECS00000041 - VITALITY/ASTRA - CUP D13C21000430001. OP is supported by the FIS2 Starting Grant FIS-2023-00246 “PhAse-sPAce cOmplexity in turbulent nearly-reversible plasmas (PAPAO)” (CUP B53C24009610001) funded by the Italian Ministry of University and Research.  
\end{acknowledgements}

\clearpage

\bibliographystyle{aa}
\bibliography{biblio}


%

\begin{appendix}
\section{The Telegraph Equation and the Quasi-Ballistic to Diffusive Transition}
\label{app:telegraph}

In this Appendix we briefly review the telegraph equation and show how it leads to the expression used in the main text for the parallel decorrelation factor $T(k_\parallel,t)$ in Equation (\ref{eq:TeleClosure}).

The telegraph equation is a hyperbolic generalization of the diffusion equation. It is particularly useful when the transport process transitions from an initial ballistic or quasi-ballistic regime to an asymptotic diffusive regime.

\subsection{From diffusion to the telegraph equation}

Let \(P(z,t)\) be the probability density of finding a particle at position \(z\) along a preferred direction, which in the present context is the direction parallel to the mean magnetic field. The standard diffusion equation is
\begin{equation}
    \frac{\partial P}{\partial t}
    =
    D_\parallel
    \frac{\partial^2 P}{\partial z^2},
    \label{eq:diffusion}
\end{equation}
where \(D_\parallel\) is the parallel diffusion coefficient, while the telegraph equation reads


\begin{equation}
    \tau_\parallel
    \frac{\partial^2 P}{\partial t^2}
    +
    \frac{\partial P}{\partial t}
    =
    D_\parallel
    \frac{\partial^2 P}{\partial z^2}.
    \label{eq:telegraph_diffusive}
\end{equation}
The telegraph equation differs from the diffusion equation in that it includes a second-order time derivative, which accounts for memory effects and finite propagation speed.

In transport problems involving particles moving with speed \(v\), one commonly writes
\begin{equation}
    D_\parallel = \frac{v^2 \tau_\parallel}{3}.
    \label{eq:Dparallel}
\end{equation}
Using Equation~\eqref{eq:Dparallel}, Equation~\eqref{eq:telegraph_diffusive} can be written as
\begin{equation}
    \frac{\partial^2 P}{\partial t^2}
    +
    \frac{1}{\tau_\parallel}
    \frac{\partial P}{\partial t}
    =
    \frac{v^2}{3}
    \frac{\partial^2 P}{\partial z^2},
    \label{eq:telegraph}
\end{equation}
where the characteristic propagation speed along the parallel direction is \(v/\sqrt{3}\) and $\tau_\parallel$ is the characteristic scattering time in the parallel direction.

\subsection{Fourier-space solution}

We now derive the expression for the Fourier-space propagator associated with Equation~\eqref{eq:telegraph}. 
We define the probability density through its Fourier transform as
\begin{equation}
    P(z,t)
    =
    \int_{-\infty}^{+\infty}
    T(k_\parallel,t)
    e^{i k_\parallel z}
    \, dk_\parallel.
    \label{eq:T_definition}
\end{equation}
where, \(T(k_\parallel,t)\) is the characteristic function of the parallel displacement,
\begin{equation}
    T(k_\parallel,t)
    =
    \left\langle
    e^{i k_\parallel \Delta z(t)}
    \right\rangle .
    \label{eq:T_characteristic}
\end{equation}


Taking the Fourier transform of Equation~\eqref{eq:telegraph}, and using
\[
    \frac{\partial^2 P}{\partial z^2}
    \longrightarrow
    - k_\parallel^2 T,
\]
One obtains the ordinary differential equation
\begin{equation}
    \frac{d^2 T}{dt^2}
    +
    \frac{1}{\tau_\parallel}
    \frac{dT}{dt}
    +
    \frac{v^2 k_\parallel^2}{3} T
    =0.
    \label{eq:T_ode}
\end{equation}
This is the equation of a damped harmonic oscillator in time.

The initial condition corresponding to a particle initially localized at the origin is
\begin{equation}
    P(z,0)=\delta(z),
\end{equation}
which implies
\begin{equation}
    T(k_\parallel,0)=1.
    \label{eq:T_initial_1}
\end{equation}
For an initially isotropic distribution of particle directions, the first time derivative of the characteristic function vanishes at \(t=0\), giving
\begin{equation}
    \left.
    \frac{dT}{dt}
    \right|_{t=0}
    =0.
    \label{eq:T_initial_2}
\end{equation}

The characteristic equation associated with Equation~\eqref{eq:T_ode} is
\begin{equation}
    s^2
    +
    \frac{1}{\tau_\parallel}s
    +
    \frac{v^2 k_\parallel^2}{3}
    =0.
\end{equation}
Its two roots are
\begin{equation}
    s_\pm
    =
    -\frac{1}{2\tau_\parallel}
    \pm
    \omega,
    \label{eq:roots}
\end{equation}
where
\begin{equation}
    \omega^2
    =
    \frac{1}{4\tau_\parallel^2}
    -
    \frac{v^2 k_\parallel^2}{3}.
    \label{eq:omega}
\end{equation}
Therefore, the general solution can be written as
\begin{equation}
    T(k_\parallel,t)
    =
    e^{-t/2\tau_\parallel}
    \left[
        A \cosh(\omega t)
        +
        B \sinh(\omega t)
    \right].
    \label{eq:T_general}
\end{equation}
The condition \(T(k_\parallel,0)=1\) gives $A=1$, while
\(dT/dt=0\) at \(t=0\) and \(A=1\), 
gives $B = \frac{1}{ 2\tau_\parallel \omega}$.

Thus,
\begin{equation}
    T(k_\parallel,t)
    =
    e^{-t/2\tau_\parallel}
    \left[
        \cosh(\omega t)
        +
        \frac{1}{2\tau_\parallel \omega}
        \sinh(\omega t)
    \right],
    \label{eq:T_solution}
\end{equation}
with \(\omega\) given by Equation~\eqref{eq:omega}. This is the expression used in the main text.

\subsection{Connection with the full displacement characteristic function}

If the perpendicular dynamics are assumed to be purely diffusive, while the telegraph equation describes the parallel dynamics, the characteristic function of the full displacement can be factorized as
\begin{equation}
    \left\langle
    e^{i \mathbf{k}\cdot \Delta \mathbf{R}(t)}
    \right\rangle
    =
    \left\langle
    e^{i \mathbf{k}_\perp\cdot \Delta \mathbf{R}_\perp(t)}
    \right\rangle
    \left\langle
    e^{i k_\parallel \Delta z(t)}
    \right\rangle.
    \label{eq:factorization}
\end{equation}
For perpendicular diffusion, one has
\begin{equation}
    \left\langle
    e^{i \mathbf{k}_\perp\cdot \Delta \mathbf{R}_\perp(t)}
    \right\rangle
    =
    e^{-D_\perp k_\perp^2 t},
    \label{eq:perp_diffusion}
\end{equation}
where \(D_\perp\) is the perpendicular diffusion coefficient. The parallel factor is precisely the telegraph solution,
\begin{equation}
    \left\langle
    e^{i k_\parallel \Delta z(t)}
    \right\rangle
    =
    T(k_\parallel,t).
\end{equation}
Therefore,
\begin{equation}
    \left\langle
    e^{i \mathbf{k}\cdot \Delta \mathbf{R}(t)}
    \right\rangle
    =
    e^{-D_\perp k_\perp^2 t}
    T(k_\parallel,t),
    \label{eq:full_characteristic}
\end{equation}
with
\begin{equation}
    T(k_\parallel,t)
    =
    e^{-t/2\tau_\parallel}
    \left[
        \cosh(\omega t)
        +
        \frac{1}{2\tau_\parallel \omega}
        \sinh(\omega t)
    \right],
\end{equation}
and
\begin{equation}
    \omega^2
    =
    \frac{1}{4\tau_\parallel^2}
    -
    \frac{v^2 k_\parallel^2}{3}.
\end{equation}

\subsection{Diffusive limit}

The telegraph equation reduces to ordinary diffusion at sufficiently long times, or equivalently at sufficiently small \(\tau_\parallel\). This can be seen directly from the roots in Equation~\eqref{eq:roots}. For
\begin{equation}
    v^2 k_\parallel^2 \tau_\parallel^2 \ll 1,
\end{equation}
we expand
\begin{equation}
    \omega
    =
    \frac{1}{2\tau_\parallel}
    \sqrt{
        1
        -
        \frac{4 v^2 k_\parallel^2 \tau_\parallel^2}{3}
    }
    \simeq
    \frac{1}{2\tau_\parallel}
    \left(
        1
        -
        \frac{2 v^2 k_\parallel^2 \tau_\parallel^2}{3}
    \right).
\end{equation}
The slow root is then
\begin{equation}
    s_+
    =
    -\frac{1}{2\tau_\parallel}
    +
    \omega
    \simeq
    -
    \frac{v^2 \tau_\parallel}{3}
    k_\parallel^2.
\end{equation}
The other root is approximately
\begin{equation}
    s_-
    \simeq
    -\frac{1}{\tau_\parallel},
\end{equation}
and therefore represents a rapidly decaying transient.

At late times, the rapidly decaying mode can be neglected, and the telegraph solution becomes
\begin{equation}
    T(k_\parallel,t)
    \simeq
    \exp
    \left(
        -D_\parallel k_\parallel^2 t
    \right),
    \label{eq:T_diffusive_limit}
\end{equation}
where
\begin{equation}
    D_\parallel
    =
    \frac{v^2 \tau_\parallel}{3}.
\end{equation}
Thus, the telegraph model recovers the usual diffusive characteristic function in the long-time limit.

\subsection{Short-time limit}

At short times, the telegraph equation retains information about the ballistic regime. Expanding $T(k_\parallel,t)$ in Taylor's series, using the initial conditions \(T(0)=1\) and \(\dot{T}(0)=0\) and Equation~(\ref{eq:T_ode}), gives
\begin{equation}
    T(k_\parallel,t)
    =
    1
    -
    \frac{v^2 k_\parallel^2}{6} t^2
    +
    O(t^3).
    \label{eq:T_short_time}
\end{equation}
This is consistent with an initially ballistic mean-square displacement,
\begin{equation}
    \left\langle
    \Delta z^2(t)
    \right\rangle
    \simeq
    \frac{v^2}{3}t^2.
\end{equation}
Therefore, the telegraph equation naturally interpolates between the quasi-ballistic regime,
\begin{equation}
    \left\langle
    \Delta z^2(t)
    \right\rangle
    \propto t^2,
    \qquad
    t \ll \tau_\parallel,
\end{equation}
and the diffusive regime,
\begin{equation}
    \left\langle
    \Delta z^2(t)
    \right\rangle
    \simeq
    2D_\parallel t,
    \qquad
    t \gg \tau_\parallel.
\end{equation}
This is the reason why Equation~\eqref{eq:T_solution} is appropriate for describing the transition between quasi-ballistic and diffusive parallel transport.

\section{Parametrization of the Parallel diffusion coefficient}
\label{app:A1}
The parallel diffusion coefficient $D_\parallel$ and the corresponding path length $\lambda_\parallel=3D_\parallel/v$ are parameterized to fit the results of our numerical simulations. The fit is given in the following form
\begin{equation}
    D_\parallel(y)=\frac{1}{3}v L_c D_0 y^\delta \left[ 1+\left(\frac{y}{R_b}\right)^{\Delta\delta/s}\right]^s,
\end{equation}
where $y=R_L/L_c$, $\delta=1/3$ and $\Delta\delta=5/3$ for a Kolmogorov spectrum in the inertial range of the turbulence, and the other parameters depend on the value of $\delta B/B_0$ and are reported in Table (\ref{tab:par}).

\begin{table}[h]
\centering
\begin{tabular}{|c|c|c|c|}
\hline
$\delta B/B_0$ & 0.3 & 0.5 & 1 \\
\hline
$D_0$ & 50.2 & 16.3 & 3.92 \\
\hline
$R_b$ & 1.92 & 1.83 & 1.92 \\
\hline
$s$ & 1.19 & 1.23 & 1.48 \\
\hline
\end{tabular}
\caption{Values of the parameters for the fit to the parallel diffusion coefficient obtained in the simulations.}
\label{tab:par}
\end{table}


\end{appendix}
\end{document}